\documentclass[aps,prb,twocolumn,a4,showpacs,superscriptaddress]{revtex4-1} 

\usepackage[english]{babel}
\usepackage[utf8x]{inputenc}
\usepackage[T1]{fontenc}
\usepackage{bm}
\usepackage{nicefrac}
\usepackage{siunitx}
\usepackage{textcomp}
\usepackage{booktabs}

\usepackage[figuresright]{rotating}

\usepackage{relsize}
\usepackage{amsmath,amssymb,amsfonts,upgreek}
\usepackage{mathtools}
\usepackage{gensymb}
\usepackage{color}
\usepackage{graphicx}
\usepackage{setspace}
\usepackage{multirow}
\usepackage{hhline}
\usepackage{bm}
\usepackage{comment}
\usepackage{natbib}
\usepackage{lipsum}

\begin{document}
\title{Understanding magnetoelectric switching in BiFeO$_3$ thin films}

\author{Natalya S.\ Fedorova}
\email{natalya.fedorova@list.lu}
\affiliation{Materials Research and Technology Department, Luxembourg Institute of Science and Technology,
5 Avenue des Hauts-Fourneaux, L-4362 Esch/Alzette, Luxembourg}

\author{Dmitri E. Nikonov}
\affiliation{Components Research, Intel Corporation, Hillsboro, 97124 Oregon, USA}

\author{John M. Mangeri}
\affiliation{Materials Research and Technology Department, Luxembourg Institute of Science and Technology,
5 Avenue des Hauts-Fourneaux, L-4362 Esch/Alzette, Luxembourg}

\author{Hai Li}
\affiliation{Components Research, Intel Corporation, Hillsboro, 97124 Oregon, USA}

\author{Ian A. Young}
\affiliation{Components Research, Intel Corporation, Hillsboro, 97124 Oregon, USA}

\author{Jorge \'{I}\~{n}iguez}
\email{jorge.iniguez@list.lu}
\affiliation{Materials Research and Technology Department, Luxembourg Institute of Science and Technology,
5 Avenue des Hauts-Fourneaux, L-4362 Esch/Alzette, Luxembourg}
\affiliation{Department of Physics and Materials Science, University of Luxembourg, 41 Rue du Brill, L-4422 Belvaux, Luxembourg}

\begin{abstract}
In this work we use a phenomenological theory of ferroelectric switching in BiFeO$_3$ thin films to uncover the mechanism of the two-step process that leads to the reversal of the weak magnetization of these materials. First, we introduce a realistic model of a BiFeO$_3$ film, including the Landau energy of isolated domains as well as the constraints that account for the presence of the substrate and the multidomain configuration found experimentally. We use this model to obtain statistical information about the switching behavior -- by running dynamical simulations based on the Landau-Khalatnikov time-evolution equation, including thermal fluctuations -- and we thus identify the factors that drive the two-step polarization reversal observed in the experiments. Additionally, we apply our model to test potential strategies for optimizing the switching characteristics.
\end{abstract}


\maketitle

\section{Introduction}
Magnetoelectric multiferroics, materials that are simultaneously magnetic and
ferroelectric, hold great potential for the development of devices with multiple functionalities and low energy consumption, as well as for their miniaturization \cite{spaldin2005,fiebig2016}. Among this class of compounds, BiFeO$_3$ is of particular interest since
it displays both ferroic orders at room temperature \cite{catalan2009}. 

Below $T_C \sim 1100$~K \cite{moreau1971,smith1968}, BiFeO$_3$ shows a spontaneous polarization $\mathbf{P}$ of up to 1 C/m$^2$ aligned along a pseudocubic $\langle
111\rangle$ direction \cite{lebeugle2007_2,wang2003} of its rhombohedrally distorted perovskite structure (space group $R3c$, \#161)
\cite{michel1969,kubel1990}. The polarization $\mathbf{P}$ originates from the displacements of Bi$^{3+}$ and Fe$^{3+}$ cations with respect to O$^{2-}$ anions, where the Bi$^{3+}$ cations dominate since they possess stereochemically active $6s$ lone
pairs \cite{seshadri2001}. 

\begin{figure*}
    \centering
    \includegraphics[width=0.95\linewidth,trim=0cm 0cm 0cm 0cm]{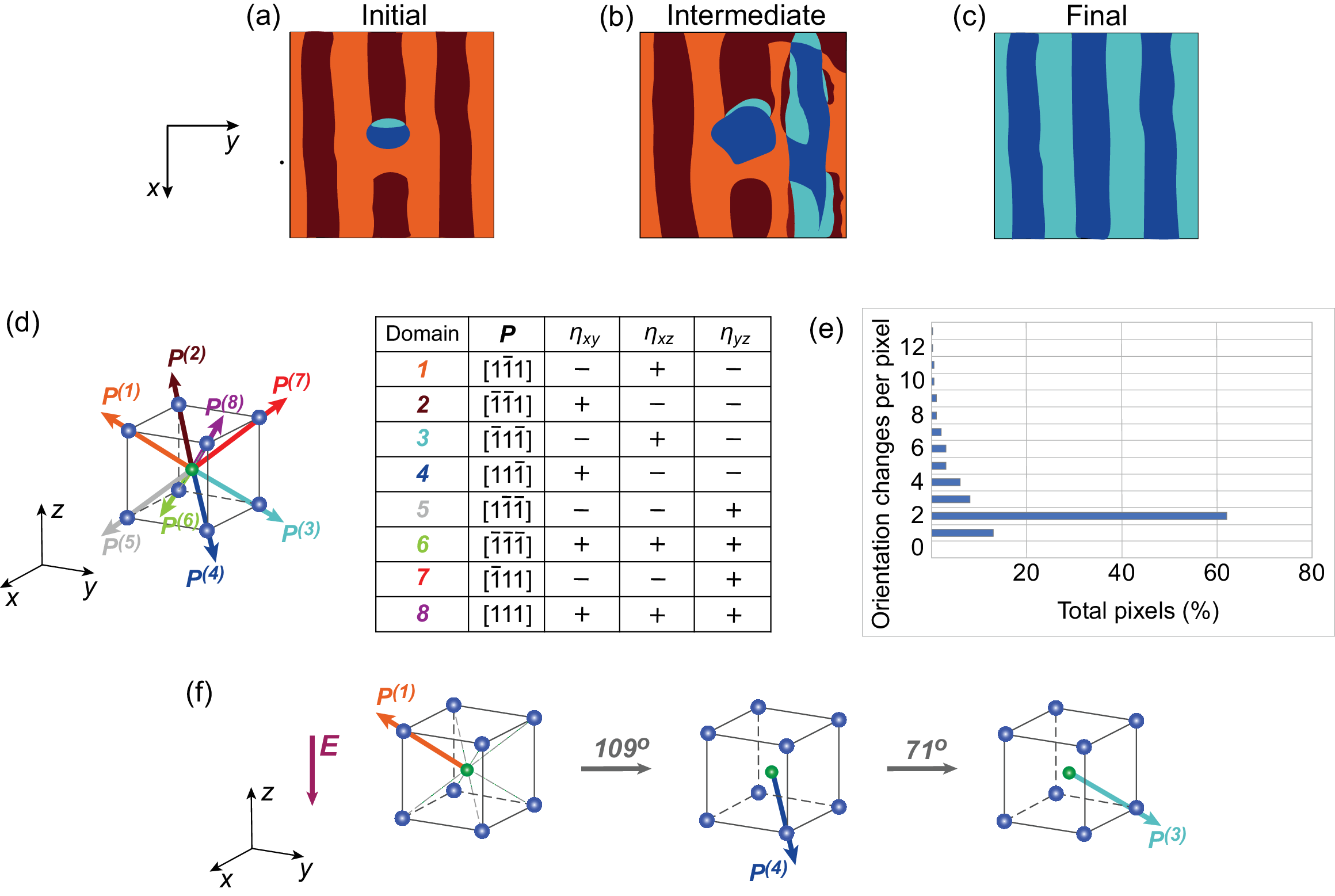}
    \caption{Polarization switching in BiFeO$_3$ films grown on a DyScO$_3$ substrate as reported by Heron \textit{et al}.~\cite{heron2014}. 
    The top row shows our hand-drawn sketches of the ferroelectric domain patterns that appear during the polarization switching process, which capture the essential features observed in the time-resolved PFM images of Refs.~\onlinecite{heron2014} and \onlinecite{private}. Panels~(a) and (c) show the ferroelectric domain patterns in the film before and after switching, respectively, while panel~(b) shows an intermediate state. Panel~(d) shows the symmetry allowed polarization variants; we also list here the $\mathbf{P}$ directions and signs of the shear strains ($\eta_{xy}$, $\eta_{yz}$ and $\eta_{xz}$) for the corresponding domains. (We use a pseudocubic setting.) Note that only the polarization variants $\mathbf{P}^{(1)}$ to $\mathbf{P}^{(4)}$ have been observed in the experiment of Heron {\it et al}. Panel~(e) shows the number of polarization switches per pixel as obtained experimentally (data taken from Ref.~\onlinecite{heron2014}). Panel~(f) illustrates the two-step polarization switching, with an electric field applied along $[00\bar{1}]$, where a $109^\circ$ $\mathbf{P}$ rotation is followed by a $71^\circ$ $\mathbf{P}$ rotation.}
\label{fig:init_final_domains} 
\end{figure*}

At $T_N \sim 640$~K, the Fe magnetic moments in BiFeO$_3$ order antiferromagnetically \cite{moreau1971,bhide1965}. The Dzyaloshinskii-Moriya (DM) interaction  \cite{dzyaloshinsky1958,moriya1960} drives the small canting of antialigned Fe spins 
which can give rise to a weak
magnetization. Crucially, the DM interaction is a consequence of the
symmetry breaking caused by the FeO$_6$ octahedral tilts of
BiFeO$_{3}$ \cite{ederer2005}. These tilts are about the axis of $\mathbf{P}$ and are expressed as $a^-a^-a^-$ in Glazer's notation \cite{glazer1972}; in the following, we denote these tilts by $\mathbf{R}$). Indeed, the phase of the octahedral
rotations defines the sign of the DM vector and, therefore, that of the
weak magnetization \cite{ederer2005}. In bulk BiFeO$_3$, an incommensurate cycloidal
spiral is superimposed on the antiferromagnetic order, which yields a zero net magnetization \cite{sosnowska1982}. However, this cycloid can be suppressed by
doping in bulk BiFeO$_3$ \cite{sosnowska2002} or by epitaxial
constraints in thin films \cite{bai2005,bea2007,sando2013,heron2014}.  In the cases of interest here, ferroelectricity coexists with weak ferromagnetism in BiFeO$_3$ at
ambient conditions \cite{wang2003,eerenstein2005,ederer2005,heron2014}. 

A deterministic reversal of the weak magnetization by an electric field was reported by Heron \textit{et al} \cite{heron2014} in a combined experimental and theoretical study of $(001)$-oriented BiFeO$_3$ films grown on a DyScO$_3$ substrate. (All the directions and plane orientations to which we refer in the text are in pseudocubic setting.) The authors demonstrated that such a magnetoelectric switching occurs as a result of a peculiar two-step polarization reversal in which a $109^\circ$ rotation of $\mathbf{P}$ is followed by a $71^\circ$ rotation (see Fig. \ref{fig:init_final_domains}(g)) or \textit{vice versa}. As it was revealed by first-principles calculations, the axis of the FeO$_6$ octahedral tilts rotates together with the polarization, which results in the reversal of the DM vector and, thus, the weak magnetization. Additionally, it was shown that a single step $180^\circ$ $\mathbf{P}$ switching, which does not affect FeO$_6$ octahedral tilts, has a significantly higher energy barrier and is therefore unfavorable. Note that such single-step $\mathbf{P}$ switching would not result in the reversal of magnetization: indeed, the observed two-step switching path is key for magnetoelectric switching to occur in multidomain BiFeO$_3$ films.

The observed possibility of manipulating a magnetization by an electric field at room temperature makes BiFeO$_3$ thim films very promising for designing novel magnetoelectric memory devices. However, to make them technologically relevant, the switching characteristics have to be optimized~\cite{manipatruni2019}.   
For example, in the experiments described above the applied voltages were in the range of a few Volts, while the current target is to switch below 100~mV; similarly, the switching times should move from the microseconds of Ref.~\onlinecite{heron2014} to values in the order of 10-1000~ps~\cite{manipatruni2018,prasad2020}. In addition, any optimization must be compatible with maintaining the magnetoelectric control, i.e., the two-step polarization switching path has to be preserved.
Therefore, to be able to optimize the switching characteristics of BiFeO$_3$ films, the microscopic mechanisms driving the two-step process need to be understood.

In this work, we introduce a phenomenological model of a multidomain BiFeO$_3$ film and investigate which physical effects (couplings) allow to reproduce the peculiar two-step polarization reversal observed experimentally. This enables us to propose potential strategies for optimizing switching characteristics, for example, via doping.  

\section{Phenomenological switching model}
\subsection{Summary of the experimental observations}
\label{subsec:experiment}

First, let us summarize the specific features of the polarization switching behavior observed in the time-resolved piezoresponce force microscopy (PFM) experiments by Heron \textit{et al.} \cite{heron2014}, which inform the definition of our phenomenological model. 

(i) $(001)$-oriented films of BiFeO$_3$ (100~nm) with SrRuO$_3$ bottom electrode (8~nm) were grown on a DyScO$_3$ substrate to obtain a striped pattern of ferroelectric domains \cite{streiffer1998,zhang2008,chu2009} with two alternating polarization directions ($\mathbf{P}^{(1)}$ and $\mathbf{P}^{(2)}$) forming $71^\circ$ domain walls as illustrated in Figs.~\ref{fig:init_final_domains}(a) and \ref{fig:init_final_domains}(d).  The electric field was applied along the $[00\bar{1}]$ direction, and it led to a $180^\circ$ polarization reversal in all domains ($\mathbf{P}^{(1)}$ switched to $\mathbf{P}^{(3)}$ and $\mathbf{P}^{(2)}$ to $\mathbf{P}^{(4)}$, following the definitions in Fig.~\ref{fig:init_final_domains}(d)). The final state is shown in Fig.~\ref{fig:init_final_domains}(c): it features a striped pattern of alternating $\mathbf{P}^{(3)}$ and $\mathbf{P}^{(4)}$ domains similar to that of the initial state (Fig.~\ref{fig:init_final_domains}(a)). Interestingly, only these four polarization variants occur during the switching process, out of the eight symmetry allowed possibilities shown in Fig.~\ref{fig:init_final_domains}(d). Also note that, in principle, one would rather expect a $71^\circ$ switching involving only the $z$ component of the polarization, which is directly coupled to the electric field applied along $[00\bar{1}]$. This switching path, however, was never observed \cite{heron2014}.

(ii) The initial and final domain configurations have the same deformation state. Indeed, all components of the strain tensor in $\mathbf{P}^{(1)}$ domains have the same magnitude and sign as those corresponding to $\mathbf{P}^{(3)}$, and the same holds for $\mathbf{P}^{(2)}$ and $\mathbf{P}^{(4)}$ domains; see Fig.~\ref{fig:init_final_domains}(d) for details.

\begin{figure}
    \centering \includegraphics[width=0.9\linewidth,trim=0cm 0cm 0cm 0cm]{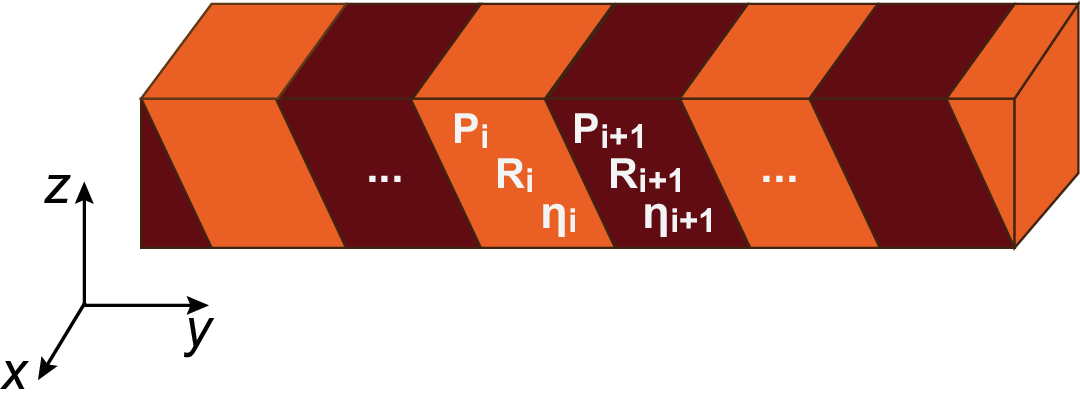}
    \caption{Approximation of a multidomain BiFeO$_3$ thin film as a one-dimensional series of domains with uniform polarization $\mathbf{P}_i$, FeO$_6$ octahedral tilts $\mathbf{R}_i$ and strain $\bm{\eta}_i$, with $i = 1, ..., N$.}
\label{fig:1dmodel} 
\end{figure}
 
(iii) For the observed polarization variants, only domain walls (DWs) in the $(011)$ plane satisfy the condition of charge neutrality, 
\begin{equation}
\sigma_{DW}=(\mathbf{P}_i-\mathbf{P}_{i+1})\mathbf{n}=0 \; ,
\label{eq:charged_DW}
\end{equation}
where $\sigma_{DW}$ is the charge density at the DW, $\mathbf{n}=(0,1,1)/\sqrt{2}$ is the vector normal to the wall plane, and $\mathbf{P}_i$ is the polarization of the $i$th domain (the domains are numbered along the $y$ direction in Fig.~\ref{fig:init_final_domains}(a), see also Fig.~\ref{fig:1dmodel}).  
Otherwise, a discontinuity in the $\mathbf{P}$ components normal to the DW plane would lead to an accumulation of bound charges, which would result in a large and unfavorable electrostatic energy penalty \cite{wu2006,dieguez2013}. 

As an example, let us consider a system of two domains with the DW in the (011) plane. If the polarization in domain \#1 is $\mathbf{P}^{(1)}=P(1,-1,1)$ and $\mathbf{P}^{(2)}=P(-1,-1,1)$ in domain \#2 (see Fig.~\ref{fig:init_final_domains}(d)), the charge density at the corresponding $71^{\circ}$~DW can be computed as (Eq.~(\ref{eq:charged_DW})): $\sigma_{DW,1-2}=P(2,0,0)\cdot(0,1,1)/\sqrt{2}=0$; therefore, this DW is neutral. Similarly, $\sigma_{DW}=0$ for $(011)$ DWs between $\mathbf{P}^{(1)}$ and $\mathbf{P}^{(3)}$ ($180^\circ$~DW), as well as between $\mathbf{P}^{(1)}$ and $\mathbf{P}^{(4)}$ ($109^\circ$~DW), and those symmetry-equivalent to them. On the other hand, the charge density at a $(011)$~DW between $\mathbf{P}^{(1)}$ and $\mathbf{P}^{(5)}$ is $\sigma_{DW,1-5}=P(0,0,2)\cdot(0,1,1)/\sqrt{2}=P\sqrt{2}$; therefore, such a DW would be charged. The same holds for $(011)$~DWs between $\mathbf{P}^{(1)}$ and $\mathbf{P}^{(6)}$, $\mathbf{P}^{(1)}$ and $\mathbf{P}^{(7)}$, and $\mathbf{P}^{(1)}$ and $\mathbf{P}^{(8)}$, and those symmetry-equivalent to them. 

We thus find the following remarkable situation. Experimentally, only the $\mathbf{P}^{(1)}$, $\mathbf{P}^{(2)}$, $\mathbf{P}^{(3)}$ and $\mathbf{P}^{(4)}$ domains have been observed during the switching, and it is known that the domain walls in the equilibrium multidomain states lie indeed within the $(011)$ plane, thus being neutral~\cite{heron2014}. This strongly suggests that the domain walls remain in the $(011)$ plane throughout the entire ferroelectric switching. This is the situation we assume in the following.

(iv) Although $180^\circ$ domain walls between $\mathbf{P}^{(1)}$ and $\mathbf{P}^{(3)}$ (or $\mathbf{P}^{(2)}$ and $\mathbf{P}^{(4)}$) would also satisfy the condition of charge neutrality described above (Eq. (\ref{eq:charged_DW})), such boundaries have never been observed experimentally. This can be seen by monitoring the domain evolution during the switching as sketched in Fig.~\ref{fig:init_final_domains}, where the orange regions never touch the light blue regions, and the brown regions are never in contact with the dark blue regions. Hence, electrostatics is not the only mechanism preventing the formation of certain boundaries. 

(v) More than 87\% of the sample area switches in more than one step. In fact, over 60\% switches in exactly two steps as shown in Fig.~\ref{fig:init_final_domains}(e). The observed steps are: a $109^\circ$ rotation of $\mathbf{P}$ where the $y$ and $z$ components are reversed (denoted as $109yz$ switch in the following) and a $71^\circ$ rotation where only the $x$ component is reversed ($71x$ switch). Interestingly, the results of Refs.~\onlinecite{heron2014,private} suggest that the $109yz$ switch is the nucleation event; this involves the reversal of the out-of-plane component, i.e., the one directly coupled to the applied field. Then, the $71x$ switches seem related to the domain wall movements and rearrangements that occur to accommodate the switched regions and eventually yield the striped state of Fig.~\ref{fig:init_final_domains}(c). Hence, in the following we assume that the $109yz$ events are the main driving force for the switching, while the $71x$ events are a secondary effect.

\subsection{Model of multidomain BiFeO$_3$ film}
\label{subsec:model}

We now introduce a model of a multidomain BiFeO$_3$ film, which will allow us to identify the minimal physical ingredients that yield the observed two-step polarization switching path. For that purpose, we take into account all the features of the switching process discussed in Sec.~\ref{subsec:experiment}. 

We approximate the BiFeO$_3$ film as a one-dimensional series of $N$ domains (Fig.~\ref{fig:1dmodel}), each of which is described by the uniform polarization $\mathbf{P}_i$, the FeO$_6$ antiphase octahedral tilts $\mathbf{R}_i$, and the strain tensor $\bm{\eta}_i$. We write the total energy density of our system as
\begin{equation}
    F=F_{L}+F_{elec}+F_{DW}+F_{elas} \; ,
    \label{eq:full_model}
\end{equation}
where $F_L$ is the self-energy of the domains (as if they were part of an infinite unconstrained bulk material); $F_{elec}$ is the electrostatic energy penalty associated to the formation of charged DWs; $F_{DW}$ is the energy penalty associated to the structural discontinuity at the DWs; and $F_{elas}$ includes the elastic constraints imposed by the DyScO$_3$ substrate as well as other elastic interactions affecting the domains. In the following, we describe these terms in detail.

\subsubsection{Energy of bulk-like domains}
\label{subsubsec:landau}

We define the self-energy of bulk-like domains as
\begin{equation}
F_{L}=\sum_{i=1}^N F_{L,i} \; ,
\label{eq:Landau}
\end{equation}
where $F_{L,i}$ is the Landau-like energy of a domain $i$ and $N$ is the total number of domains considered in our simulations. 

Following Ref.~\onlinecite{fedorova2022}, we represent $F_{L,i}$ as an expansion around the paraelectric cubic perovskite structure in powers of the electric polarization $\mathbf{P}_i=(P_{i,x},P_{i,y},P_{i,z})$, the FeO$_6$ antiphase octahedral tilts $\mathbf{R}_i=(R_{i,x},R_{i,y},R_{i,z})$ and the strain $\bm{\eta}_i=(\eta_{i,xx},\eta_{i,yy},\eta_{i,zz},\eta_{i,yz},\eta_{i,xz},\eta_{i,xy})$. Here 
$\eta_{i,xx}=\epsilon_{i,xx}$, $\eta_{i,yy}=\epsilon_{i,yy}$,
$\eta_{i,zz}=\epsilon_{i,zz}$, $\eta_{i,yz}=2\epsilon_{i,yz}$,
$\eta_{i,xz}=2\epsilon_{i,xz}$ and $\eta_{i,xy}=2\epsilon_{i,xy}$, and
$\epsilon_{i,\alpha\beta}$ are the symmetric components of the homogeneous strain
tensor. The resulting expression for the potential is written as
\begin{equation}
\label{eq:E_Landau_single}
\begin{aligned}
    F_{L,i}=&F_0+F(\mathbf{P}_i)+F(\mathbf{R}_i)+F(\bm{\eta}_i)+ \\ 
    & F(\mathbf{P}_i,\mathbf{R}_i)+F(\mathbf{P}_i,\bm{\eta}_i)+F(\mathbf{R}_i,\bm{\eta}_i) \; ,
    \end{aligned}
\end{equation}
where $F_0$ is the energy of the reference cubic phase of bulk BiFeO$_3$.
$F(\mathbf{P}_i)$, $F(\mathbf{R}_i)$, and $F(\bm{\eta}_i)$ are the energy contributions due to polarization, FeO$_6$ tilts and strain, respectively, and have the following expressions: 
\begin{equation}
\label{eq:F_P}
\begin{aligned}
    F(\mathbf{P}_i)=&A_P(P_{i,x}^2+P_{i,y}^2+P_{i,z}^2)+ \\ 
    & B_P(P_{i,x}^2+P_{i,y}^2+P_{i,z}^2)^2+\\
    & C_P(P_{i,x}^2P_{i,y}^2+P_{i,y}^2P_{i,z}^2+P_{i,z}^2P_{i,x}^2) \; ,
\end{aligned}
\end{equation}
\begin{equation}
\label{eq:F_R}
\begin{aligned}
    F(\mathbf{R}_i)=&A_R(R_{i,x}^2+R_{i,y}^2+R_{i,z}^2)+ \\ 
    & B_R(R_{i,x}^2+R_{i,y}^2+R_{i,z}^2)^2+\\
    & C_R(R_{i,x}^2R_{i,y}^2+R_{i,y}^2R_{i,z}^2+R_{i,z}^2R_{i,x}^2) \; , 
\end{aligned}
\end{equation}
and
\begin{equation}
\label{eq:F_eta}
\begin{aligned}
    F(\bm{\eta}_i)=& \frac{1}{2}C_{11}(\eta_{i,xx}^2+\eta_{i,yy}^2+\eta_{i,zz}^2)+ \\ 
    & C_{12}(\eta_{i,xx}\eta_{i,yy}+ \eta_{i,yy}\eta_{i,zz}+\eta_{i,zz}\eta_{i,xx})+ \\ & \frac{1}{2}C_{44}(\eta_{i,yz}^2+\eta_{i,xz}^2+\eta_{i,xy}^2).
\end{aligned}
\end{equation}
Since our goal is to introduce a minimal model that captures the basic energetics of BiFeO$_3$, we truncate the expansion in $\mathbf{P}_i$ and $\mathbf{R}_i$ at the fourth
order (the minimum necessary to model structural
instabilities) and only include harmonic terms for $\bm{\eta}_i$. Then, 
$F(\mathbf{P}_i,\mathbf{R}_i)$, $F(\mathbf{P}_i,\bm{\eta}_i)$ and $F(\mathbf{R}_i,\bm{\eta}_i)$ are the coupling terms between the considered degrees of freedom:
\begin{equation}
\label{eq:F_PR}
    \begin{aligned}
    F(&\mathbf{P}_i,\mathbf{R}_i)=B_{PR}(P_{i,x}^2+P_{i,y}^2+P_{i,z}^2)(R_{i,x,}^2+R_{i,y}^2+R_{i,z}^2)+ \\
    & C_{PR}(P_{i,x}^2R_{i,x}^2+P_{i,y}^2R_{i,y}^2+P_{i,z}^2R_{i,z}^2)+ \\
    & C'_{PR}(P_{i,x} P_{i,y} R_{i,x} R_{i,y}+P_{i,y} P_{i,z} R_{i,y} R_{i,z}+ \\ 
    & P_{i,z} P_{i,x} R_{i,z} R_{i,x}) \; ,
    \end{aligned}
\end{equation}
\begin{equation}
\label{eq:F_Peta}
    \begin{aligned}
    &F(\mathbf{P}_i,\bm{\eta}_i)= \gamma_{P111}(\eta_{i,xx}P_{i,x}^2+\eta_{i,yy}P_{i,y}^2+\eta_{i,zz}P_{i,z}^2) + \\ & \gamma_{P122}(\eta_{i,xx}(P_{i,y}^2+P_{i,z}^2)+ \eta_{yy}(P_{i,z}^2+P_{i,x}^2)+ \\ & \eta_{i,zz}(P_{i,x}^2+P_{i,y}^2))+\\ & \gamma_{P423}(\eta_{i,yz}P_{i,y}P_{i,z}+\eta_{i,xz}P_{i,z}P_{i,x}+\eta_{i,xy}P_{i,x}P_{i,y}) \; ,  
    \end{aligned}
\end{equation}
and
\begin{equation}
\label{eq:F_Reta}
    \begin{aligned}
    &F(\mathbf{R}_i,\bm{\eta}_i)= \gamma_{R111}(\eta_{i,xx}R_{i,x}^2+\eta_{i,yy}R_{i,y}^2+\eta_{i,zz}R_{i,z}^2) + \\ & \gamma_{R122}(\eta_{i,xx}(R_{i,y}^2+R_{i,z}^2)+ \eta_{i,yy}(R_{i,z}^2+R_{i,x}^2)+ \\ & \eta_{i,zz}(R_{i,x}^2+R_{i,y}^2))+\\ & \gamma_{R423}(\eta_{i,yz}R_{i,y}R_{i,z}+\eta_{i,xz}R_{i,z}R_{i,x}+\eta_{i,xy}R_{i,x}R_{i,y}) \; .
    \end{aligned}
\end{equation}
Again, for simplicity we only include in the model the lowest-order symmetry-allowed couplings.

In Eqs.~(\ref{eq:F_P})-(\ref{eq:F_Reta}), $A$, $B$, $C$, $C'$ and $\gamma$ are the 
expansion coefficients that one can compute for bulk BiFeO$_3$ using density functional theory (DFT). (For convenience, the coefficients $C_{11}$, $C_{12}$ and $C_{44}$ of Eq.~(\ref{eq:F_eta}) as well as $\gamma$ of Eqs.~(\ref{eq:F_Peta}) and (\ref{eq:F_Reta}) are given in Voigt notation.) The values are given in Table~\ref{tab:default_set}. This model allows to accurately predict the energies and structural properties of BiFeO$_3$ polymorphs that are relevant for the purposes of studying polarization switching. Additional details on the model, the physical insights that it provides, as well as the methods for computing its parameters can be found in Ref.~\onlinecite{fedorova2022}. 

\subsubsection{Electrostatic energy penalty}
\label{subsubsec:electrostatic}

The formation of charged domain walls gives rise to a large and unfavorable electrostatic energy penalty, which makes them unstable (in the absence of charged defects or free carriers). Therefore, we need to ensure that our model precludes the formation of such boundaries. For this purpose, we introduce the term
\begin{equation}
    F_{elec}=\frac{1}{2} K_{elec} \sum_{i=1}^{N-1} \lvert \left(\mathbf{P}_i-\mathbf{P}_{i+1}\right)\cdot \mathbf{n} \rvert^2,
    \label{eq:F_elec}
\end{equation}
where $K_{elec}$ is an electrostatic penalty constant. Calculating $K_{elec}$ from first principles would require DFT simulations of charged domain walls in BiFeO$_3$. This, however, is very challenging due to their instability. Thus, since charged domain walls are not observed during polarization switching in the experiments of interest here, we simply assume $K_{elec}>0$ with a magnitude large enough to prevent their formation in our simulations (see Table~\ref{tab:default_set}).  

\subsubsection{Structural energy of domain walls}

The next term, $F_{DW}$, describes the energy penalty due to structural discontinuity associated to the formation of DWs, and we define it as
\begin{equation}
    F_{DW}=F_{DW,P}+F_{DW,R} \; .
    \label{eq:F_DW_full}
\end{equation}
Here, the $F_{DW,P}$ contribution arises from the change of the polarization across the wall, and we write it as
\begin{equation}
    F_{DW,P}=\frac{1}{2} K_{DW,P} \sum_{i=1}^{N-1} \sum_{\alpha=x,y,z} \left(P_{i,\alpha}-P_{i+1,\alpha}\right)^2 \; .
\label{eq:F_DW_P}
\end{equation}
In turn, the $F_{DW,R}$ term penalizes a discontinuity in the octahedral tilt pattern, and has the form
\begin{equation}
    F_{DW,R}=\frac{1}{2} K_{DW,R} \sum_{i=1}^{N-1} \sum_{\alpha=x,y,z} \left(R_{i,\alpha}-R_{i+1,\alpha}\right)^2 \; .
\label{eq:F_DW_R}
\end{equation}
As one can see from Eqs.~(\ref{eq:F_DW_P}) and (\ref{eq:F_DW_R}), the domain wall energy is minimized if polarization and octahedral tilts are 
uniform across the sample (i.e., for a monodomain configuration).   We use the DFT domain wall energies computed for BiFeO$_3$ by Dieguez \textit{et al}. \cite{dieguez2013} to  estimate the parameters $K_{DW,P}$ and $K_{DW,R}$. Note that the obtained values (Table~\ref{tab:default_set}) reflect the fact that the main contribution to $F_{DW}$ is given by $F_{DW,R}$ (see more details in Sec.~SI of the Supplementary material). 

\begingroup
\setlength{\tabcolsep}{9pt} 
\renewcommand{\arraystretch}{1.2} 
\begin{table}
\caption{Our initial guess ("default set")
for the parameters of the model introduced in Sec.~\ref{subsec:model} for multidomain BiFeO$_3$ thin films. The parameters of the Landau model ($A_P$, $A_R$, $B_P$, ...) \cite{fedorova2022} as well as $K_{elec}$, $K_{elas,xy}$, $K_{elas,xz}$ and $K_{elas,yz}$, are normalized so they give  energies per 5-atom perovskite unit cell. The $K_{DW,P}$ and $K_{DW,R}$ parameters give domain wall energies per unit area (see Sec.~SI of the Supplementary Material for details on how we obtain them).}
\begin{tabular}{ccc}
\hline
\hline
& \centering{Default set} & Units
\tabularnewline
\hline
\centering{$A_P$} & -1.747 & \centering{$\times10^{-19}$, J m$^4$ C$^{-2}$}
\tabularnewline
\centering{$B_P$} & 1.070 & \centering{$\times10^{-19}$, J m$^8$ C$^{-4}$}
\tabularnewline
\centering{$C_P$} & -7.486 & \centering{$\times10^{-20}$, J m$^8$ C$^{-4}$}
\tabularnewline
\centering{$A_R$} & -8.555  & \centering{$\times10^{-22}$, J  deg$^{-2}$}
\tabularnewline
\centering{$B_R$} & 2.169 & \centering{$\times10^{-24}$, J  deg$^{-4}$}
\tabularnewline
\centering{$C_R$} & -1.240 & \centering{$\times10^{-24}$, J deg$^{-4}$}
\tabularnewline
\centering{$C_{11}$} & 1.833 & \centering{$\times10^{-17}$, J}
\tabularnewline
\centering{$C_{12}$} & 7.301 &  \centering{$\times10^{-18}$, J}
\tabularnewline
\centering{$C_{44}$} & 4.600 & \centering{$\times10^{-18}$, J}
\tabularnewline
\centering{$B_{PR}$} &  1.121 & \centering{$\times10^{-21}$, J m$^4$ C$^{-2}$ deg$^{-2}$}
\tabularnewline
\centering{$C_{PR}$} & -3.437 & \centering{$\times10^{-22}$, J m$^4$ C$^{-2}$ deg$^{-2}$}
\tabularnewline
\centering{$C'_{PR}$} & -2.245 & \centering{$\times10^{-21}$, J m$^4$ C$^{-2}$ deg$^{-2}$}
\tabularnewline
\centering{$\gamma_{P111}$} & -9.444 & \centering{$\times10^{-19}$, J m$^4$ C$^{-2}$}
\tabularnewline
\centering{$\gamma_{P122}$} & -1.557 & \centering{$\times10^{-19}$, J m$^4$ C$^{-2}$}
\tabularnewline
\centering{$\gamma_{P423}$} & -3.232 & \centering{$\times10^{-19}$, J m$^4$ C$^{-2}$}
\tabularnewline
\centering{$\gamma_{R111}$} & -1.178 &  \centering{$\times10^{-21}$, J  deg$^{-2}$}
\tabularnewline
\centering{$\gamma_{R122}$} & 1.158 & \centering{$\times10^{-22}$, J   deg$^{-2}$}
\tabularnewline
\centering{$\gamma_{R423}$} & 1.155 & \centering{$\times10^{-21}$, J   deg$^{-2}$}
\tabularnewline
\hline
\centering{$K_{elec}$} & 1.602 & \centering{$\times 10^{-19}$, J m$^4$ C$^{-2}$}
\tabularnewline
\centering{$K_{DW,P}$} & 4.756 & \centering{$\times 10^{-2}$, J m$^2$ C$^{-2}$}
\tabularnewline
\centering{$K_{DW,R}$} & 4.733 & \centering{$\times 10^{-4}$, J m$^{-2}$ deg$^{-2}$}
\tabularnewline
\centering{$K_{elas,xy}$} & 4.600 & \centering{$\times 10^{-18}$, J}
\tabularnewline
\centering{$K_{elas,xz}$} & 4.600 & \centering{$\times 10^{-18}$, J}
\tabularnewline
\centering{$K_{elas,yz}$} & 4.600 & \centering{$\times 10^{-18}$, J}
\tabularnewline
\hline
\centering{$L_P$} & 2.000 & \centering{$\times10^2$ F m$^{-1}$ s$^{-1}$}
\tabularnewline
\centering{$L_R$} & 8.319 & \centering{$\times10^4$ deg$^2$ m$^3$ J$^{-1}$ s$^{-1}$}
\tabularnewline
\centering{$L_\eta$} & 1.714 & \centering{m$^3$ J$^{-1}$ s$^{-1}$}
\tabularnewline
\hline
\centering{$Q_P$} & 2.000 & \centering{$\times 10^2$ F m$^{-1}$ s$^{-1}$}
\tabularnewline
\centering{$Q_R$} & 8.319 & \centering{$\times 10^4$ deg$^2$ m$^3$ J$^{-1}$ s$^{-1}$}
\tabularnewline
\hline
\hline
\end{tabular}
\label{tab:default_set}
\end{table}

\subsubsection{Elastic constraints}
\label{subsubsec:elastic_constraints}

We describe the elastic constraints imposed on the system as
\begin{equation}
F_{elas}=F_{sub}+F_{matrix} \; .
\label{eq:F_elas_full}
\end{equation}
Here, $F_{sub}$ takes into account the action of the DyScO$_3$ substrate, which forces the BiFeO$_3$ film to have no net shear strain in the $xy$ plane. This leads to the formation of the experimentally observed pattern of ferroelectric {\sl and ferroelastic} domains (with alternating $\mathbf{P}^{(1)}/\mathbf{P}^{(2)}$ polarization variants in the initial state, and $\mathbf{P}^{(3)}/\mathbf{P}^{(4)}$  variants in the final state) where the shear $\eta_{i,xy}$ alternate sign between neighboring domains \cite{winchester2011} (see Figs.~\ref{fig:init_final_domains}(a), \ref{fig:init_final_domains}(c) and \ref{fig:init_final_domains}(d)). To favor the observed ground state, we write $F_{sub}$ as
\begin{equation}
    F_{sub}=\frac{1}{2}K_{elas,xy}\sum_{i=1}^{N-1}\left(\eta_{i,xy}+\eta_{i+1,xy}\right)^2 \; ,
    \label{eq:F_sub}
\end{equation}
 where $K_{elas,xy}>0$. 

Additionally, as we mentioned in Sec.~\ref{subsec:experiment}, it is experimentally found that the deformation state is preserved during the switching process, i.e., the domains in the initial and final state have the exact same strains. Following the discussion by Heron \textit{et al}.~\cite{heron2014}, we hypothesize that this may result from the interaction between switching domains and the matrix of as-grown domains that are not yet switched, since the electric field is applied locally in the discussed experiments. To take this into account, we write $F_{matrix}$ as
\begin{equation}
    F_{matrix}=F_{elas,xz}+F_{elas,yz} \; ,
    \label{eq:F_matrix}
\end{equation}
where
\begin{equation}
    F_{elas,xz}=\frac{1}{2} K_{elas,xz}\sum_{i=1}^N \left(\eta_{i,xz}-\eta_{i,xz}^{ref}\right)^2
    \label{eq:F_elas_xz}
\end{equation}
and
\begin{equation}
    F_{elas,yz}=\frac{1}{2} K_{elas,yz}\sum_{i=1}^N \left(\eta_{i,yz}-\eta_{yz}^{ref}\right)^2 \; .
    \label{eq:F_elas_yz}
\end{equation}
Here, $\eta_{i,xz}^{ref}>0$ ($\eta_{i,xz}^{ref}<0$) for odd $i$ (for even $i$) are the strains corresponding to the initial state or the unswitched region (see Fig.~\ref{fig:init_final_domains}(d)). In turn, $\eta_{yz}^{ref}<0$ for all observed polarization variants ($\mathbf{P}^{(1)}$ to $\mathbf{P}^{(4)}$).  The magnitude of $\eta_{i,xz}^{ref}$ and $\eta_{yz}^{ref}$ is obtained by minimizing the energy of single domain BiFeO$_3$ (Eq.~\ref{eq:E_Landau_single}) with no applied electric field ; we get $|\eta_{i,xz}^{ref}|=|\eta_{yz}^{ref}|=0.0059$.

As an initial guess, we take $K_{elas,xy}$, $K_{elas,xz}$ and $K_{elas,yz}$ equal to the elastic constant $C_{44}$ as computed for BiFeO$_3$ using DFT, which captures the stiffness against a shear strain that is typical of perovskite oxides (see Table~\ref{tab:default_set}). Note that this is a very crude approximation. For example, the energy penalty associated to the strain relaxation of the BiFeO$_3$ film (so it adopts its preferred shear strain and overcomes the clamping by the substrate) will be associated to the formation of misfit dislocations \cite{hull1992}. By contrast, the energy penalty imposed by a typical $C_{44}$ can be expected to be much smaller. Hence, in the absence of a better quantitative guideline, the values of $K_{elas}$ constants ($K_{elas,xy}$ in particular) will play the role of adjustable parameters in our discussion.

\subsubsection{Effect of domain size}
\label{subsubsec:domain_size}

In the sections above we have introduced all the terms in our model potential $F$. Now, to construct the total energy of a particular multidomain structure, we have to pay attention to its specific dimensions (e.g., its domain width) and how the different energy terms scale with them. 

For example, the Landau energy $F_{L,i}$ and the elastic energy penalty $F_{elas}$ scale with the domain width. By contrast, the domain wall energy $F_{DW}$ is defined by the discontinuity in $\mathbf{P}$ and $\mathbf{R}$ at the wall and it is therefore independent of the domain width. Finally, the electrostatic penalty $F_{elec}$ scales with the domain width; yet, this is not important here, since we choose $K_{elec}$ large enough so that the formation of charged domain boundaries is fully precluded. 

Thus, following experiments \cite{shelke2012,heron2014}, let us assume that our state of interest features domains that are 100~nm wide, which is typical of BiFeO$_{3}$ films about 100~nm thick. For simplicity, we use a domain volume $V_{d}=$(100 nm)$^3$, noting that the length of the third dimension plays no role in the problem. This would correspond to domains with about $N_{uc,d}=V_{d}/V_{uc} = 1.7\times 10^{7}$ 5-atom BiFeO$_{3}$ unit cells in them, using $V_{uc}=(0.3866$~nm$)^3$ for the cell volume. Similarly, we have DWs with an area $S_{DW} = (100$~nm$)^{2}$.

Now, in Table~\ref{tab:default_set} we give the parameters $A_P$, $B_P$, $C_P$,..., $\gamma_{R423}$ to compute the domain self-energy {\sl per unit cell (uc)}, which we denote $F_{L,uc}$. By contrast, the parameters $K_{DW,P}$ and $K_{DW,R}$ in Table~\ref{tab:default_set} allow us to compute the DW energy {\sl per unit area (ua)} associated to a discontinuity in polarization or tilts, $F_{DW,ua}$. Hence, to compute the total energy, we have to weight these quantities appropriately as $N_{uc,d}F_{L,uc} + S_{DW}F_{DW,ua}$. More generally, we work with an energy density $f$ (normalized by $V_d$) as
\begin{widetext}
\begin{equation}
\begin{aligned}
    f & =\frac{F}{V_d} - {\bf E} \sum_i {\bf P}_i =\frac{1}{V_{d}}(F_{L}+F_{elec}+F_{elas}+F_{DW})  - {\bf E} \sum_i {\bf P}_i \\
    & =\frac{1}{V_{uc}N_{uc,d}} (N_{uc,d}F_{L,uc}+N_{uc,d}F_{elec,uc}+N_{uc,d}F_{elas,uc} +S_{DW}F_{DW,ua})  - {\bf E} \sum_i {\bf P}_i \\
    & =\frac{1}{V_{uc}} (F_{L,uc}+F_{elec,uc}+F_{elas,uc}) +\frac{1}{V_{d}} S_{DW}F_{DW,ua}  - {\bf E} \sum_i {\bf P}_i \; ,
    \end{aligned}
    \label{eq:energy_density}
\end{equation}
\end{widetext}
where we introduce various unit-cell normalized energies and ${\bf E}$ is an external electric field.
 
\subsection{Numerical approach to finite temperature dynamics}
\label{subsec:finite_temp_dynamics}

In order to study the switching dynamics, we solve a system of Landau-Khalatnikov time-evolution equations (LKEs) of the form\cite{umantsev2012}
\begin{equation}
    \frac{d\phi_{i}}{dt}=-L_{\phi} \frac{\partial f}{\partial \phi_i}+\theta_{\phi_{i}}(t) \; ,
\label{eq:LK}
\end{equation}
where $\phi_{i}$ denotes the order parameters ($\mathbf{P}_i$, $\mathbf{R}_i$ or $\bm{\eta}_i$) in domain $i$; $t$ is the time; $f$ is the energy density of Eq.~(\ref{eq:energy_density}); and $L_\phi$ is the kinetic coefficient that defines the rate at which an order parameter $\phi$ approaches its equilibrium value. The latter are chosen such that the relaxation times for all order parameters are in the range of picoseconds (see Table~\ref{tab:default_set}). Since we are interested in studying polarization switching dynamics at room temperature, we introduce a stochastic noise  
$\theta_{\phi_{i}}(t)$ for $\phi_i=\mathbf{P}_i$ and $\mathbf{R}_i$ that resembles thermal fluctuations of $\phi_i$ following Ref.~\onlinecite{Indergand2020}. $\theta_{\phi_i}(t)$ obeys a Gaussian probability distribution and is uncorrelated in time. Its auto-correlation function is:
\begin{equation}
    \langle \theta_{\phi_{i}}(t) \theta_{\phi_{i}}(t')\rangle=\frac{2k_B T Q_\phi}{V_{uc} \Delta t} \delta(t-t') \; ,
\end{equation}
where $k_B$ is Boltzmann's constant, $\Delta t$ is the time step used in the simulations, $Q_\phi$ defines the noise amplitude and is often chosen to be equal to the kinetic coefficient $L_\phi$.  Hence, at each time step in the simulations we add the noise given by
\begin{equation}
    \theta_{\phi_{i}}=\sqrt{\frac{2k_B T Q_\phi}{V_{uc} \Delta t}}n \; ,
    \label{eq:noise_term}
\end{equation}
where $n$ is a random number out of a normal distribution $N(\mu,\sigma^2)$ with mean $\mu=0$ and variance $\sigma^2=1$.  

Additionally, to account for temperature-driven changes in the free energy landscape, 
we include a temperature dependence in the self-energy of the domains, Eq.~(\ref{eq:E_Landau_single}). For simplicity, we modify only the terms/expressions which presumably have the strongest impact on the materials response properties.  More specifically, instead of using constant values for the coefficients $A_P$ and $A_R$ entering $F(\mathbf{P}_i)$ and $F(\mathbf{R}_i)$ (Eqs.~(\ref{eq:F_P}) and (\ref{eq:F_R}), respectively), we define them as follows  \cite{devonshire1954,rabe2007}:
\begin{equation}
    A_\phi=A_\phi^{(0)}\frac{T_{C,\phi}-T}{T_{C,\phi}} \; ,
    \label{eq:A_temp_dependence}
\end{equation}
where $\phi$ stands for $\mathbf{P}$ or $\mathbf{R}$; $A_\phi^{(0)}<0$ is the corresponding coefficient as obtained using DFT (thus, at $T=0$~K, see Table~ \ref{tab:default_set}); $T_{C,\phi}$ is the critical temperature below which $\phi$ appears in BiFeO$_3$ (ignoring the subtleties of the structural phase transitions in this material \cite{catalan2009}, we simply choose $T_{C,P}=1143$ K and $T_{C,R}=1205$ K); and $T$ is the operating temperature of interest ($T=300$~K). 

\subsection{Adjustable parameters} 
\label{subsubsec:calibration}

As we discussed in Sec.~\ref{subsec:model}, we can estimate most of the parameters of our model using DFT results. In the following, we will refer to this parameter set as the ``default set'' (the values are given in Table~\ref{tab:default_set}).  However, we need to ensure that our model reproduces the experimentally observed switching behavior of BiFeO$_3$ films at room temperature.  

In particular, the Landau model parameters define the depth of potential wells corresponding to the states with different polarization directions and senses of the FeO$_6$ octahedral rotations, as well as the height of the energy barriers between these states. Therefore, these parameters determine coercive fields and switching times in our simulations.  
Although we introduce a finite temperature correction to the quadratic coefficients $A_P$ and $A_R$ of the Landau potential (Eqs.~(\ref{eq:F_P}) and (\ref{eq:F_R}), respectively, as well as Eq.~(\ref{eq:A_temp_dependence})), this might not be sufficient to reproduce the experimental switching characteristics of BiFeO$_3$ films. For example, we noticed that our DFT values of the Landau model parameters are generally one order of magnitude larger than those typically used in phase-field simulations of BiFeO$_3$ at room temperature~\cite{xue2014,shi2022}. Since the latter allow operating with switching voltages that are close to the experimental values, this suggests that our DFT model parameters  overestimate the height of the energy barriers.  Moreover, our model does not describe the presence of defects that serve as nucleation centers and reduce energy barriers for polarization switching. Therefore, to take these effects into account, we introduce an additional parameter $K$ in the range from 0.6 to 1 that serves as a re-scaling factor for the Landau energy $F_L$ (Eq.~(\ref{eq:E_Landau_single})) and the domain wall energy $F_{DW}$ (Eq.~(\ref{eq:F_DW_full})). For the elastic constraint terms, we consider several values of $K_{elas,xy}=K_{elas,xz}$ while keeping $K_{elas,yz}$ (Eq.~(\ref{eq:F_elas_yz})) fixed to the initial DFT estimate since it only penalizes the states with polarization $\mathbf{P}^{(5)}$ to $\mathbf{P}^{(8)}$ (see Fig.~\ref{fig:init_final_domains}) and, as we will demonstrate below, does not play a significant role in driving two-step polarization switching process.  

Additionally, as we discussed in Sec.~\ref{subsec:finite_temp_dynamics}, we take into account thermal fluctuations of the order parameters by adding the stochastic noise term (Eq.~(\ref{eq:noise_term})) in the LKEs (Eq.~(\ref{eq:LK})). Overall, thermal fluctuations allow to overcome switching energy barriers more easily, and therefore should reduce switching fields and times in our simulations. As an initial guess for the noise amplitudes, we choose $Q_\phi=L_\phi$ (default parameter set in Table~\ref{tab:default_set}). One has to keep in mind, however, 
that if these noise amplitudes are too large, we may obtain spurious switches of $\mathbf{P}$ and $\mathbf{R}$, which are not observed in experiments. On the other hand, as we discuss below, too small $Q_\phi$ values may make $71x$ switching events too unlikely, as these are not directly driven by the applied electric field along $[00\bar{1}]$ direction.
Therefore, we consider multiple combinations of $Q_P$ and $Q_R$ and select those giving the best agreement with the experiment (details below, in Sec.~\ref{subsec:validation}).  

\subsection{Calculation details}

We perform polarization switching simulations based on the model presented in Sec.~\ref{subsec:model} and LKEs introduced in Sec.~\ref{subsec:finite_temp_dynamics} using an in-house developed code. In all simulations, the kinetic coefficients $L_\phi$ of LKEs are set equal to the values in Table~\ref{tab:default_set}. 
We numerically solve the system of LKEs using the classic Runge-Kutta method~\cite{butcher2016}. We use a constant time step $\delta t=40$~fs and run each simulation for $t_{tot}=280$~ns. 
Unless otherwise specified, for each choice of model and simulation parameters, we perform 200 runs with different random noise, to obtain statistics of the switching events. 

We apply a sinusoidal electric field $E(t)=E_{max} \sin( 2\omega t)$ along $[00\bar{1}]$. Following the experiments of Heron \textit{et al}.~\cite{heron2014}, we use $\omega=1.8$~MHz. $E_{max}$ values are taken in the range of 100-700~MV/m as will be detailed below.   

\section{Results}
\subsection{Validation of the model}
\label{subsec:validation}

\subsubsection{Polarization switching in  monodomain BiFeO$_3$}
\label{subsubsec:monodomain}

\begin{figure}
    \centering \includegraphics[width=0.98\linewidth,trim=0cm 0cm 0cm 0cm]{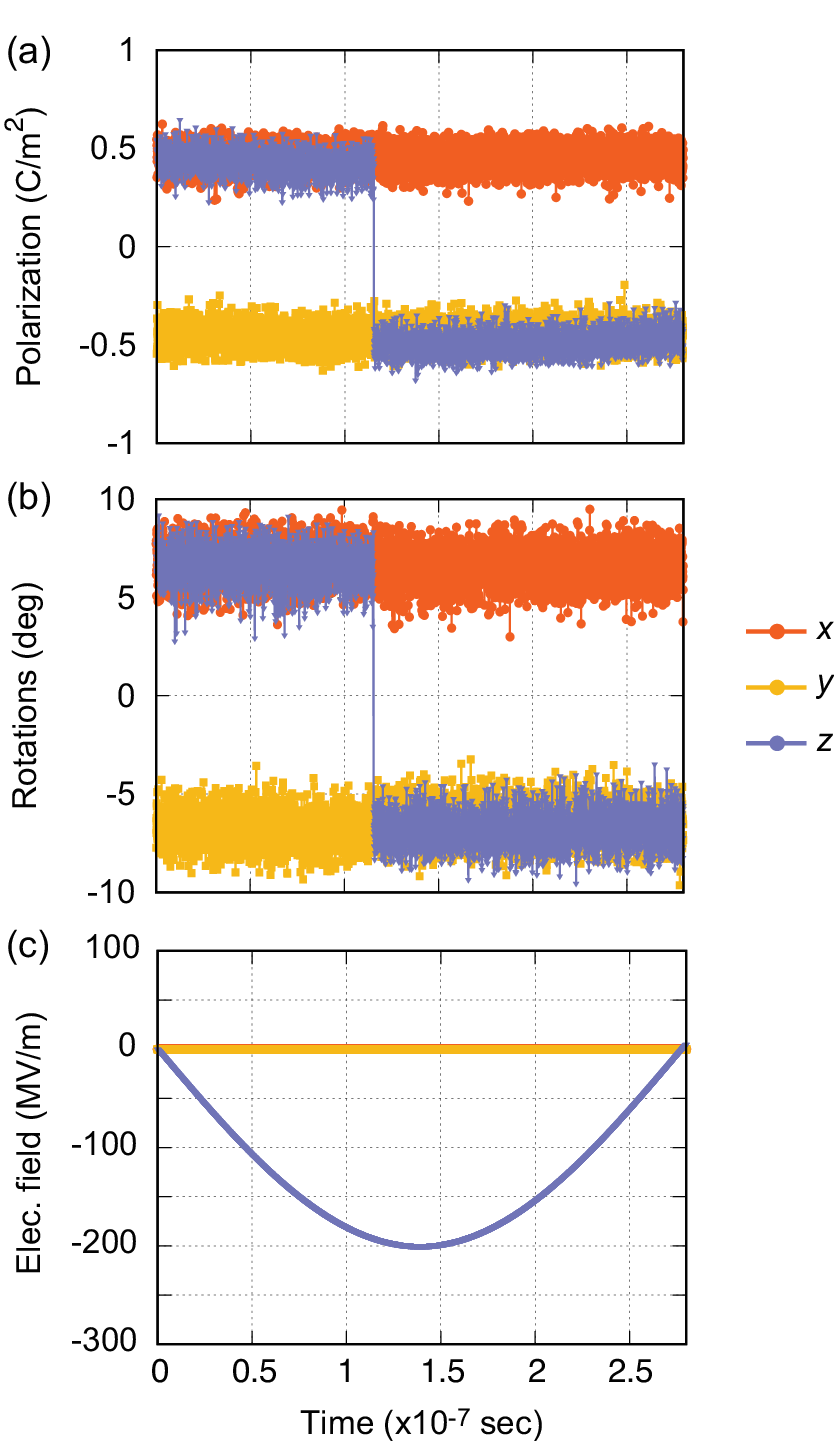}
    \caption{Time evolution of the order parameters in monodomain bulk BiFeO$_3$ simulated at $T=300$~K. We show the evolution of the polarization (a), the FeO$_6$ octahedral tilts (b), and the applied electric field (c).}
\label{fig:switching_mono} 
\end{figure}

First, we study the simple case of monodomain bulk BiFeO$_3$.
The energy of this system is described solely by the Landau energy term, Eq.~(\ref{eq:E_Landau_single}). Here we use the default values of the Landau model parameters presented in Table~\ref{tab:default_set}. Our starting configuration (before switching) is shown in Fig.~\ref{fig:switching_behavior}(a) with $\mathbf{P}$, $\mathbf{R}$ and $\bm{\eta}$ obtained by minimizing the energy of the system at $T=0$~K with no applied electric field.

We start by identifying suitable noise amplitudes for $\mathbf{P}$ and $\mathbf{R}$ ($Q_P$ and $Q_R$, respectively). For that purpose, we analyze time evolution of the order parameters in this system at $E=0$~MV/m and $T=300$~K with different $Q_P$ and $Q_R$. In such conditions, the order parameters are expected to fluctuate around their initial state, but no (spontaneous) switching events should occur.  We begin with the default values of the noise amplitudes, namely, $Q_P=L_P=200$~F~m$^{-1}$~s$^{-1}$ and $Q_R=L_R=8.319\times10^4$~deg$^2$~m$^3$ J$^{-1}$~s$^{-1}$. We repeated this simulation keeping the same conditions 200 times in order to obtain the statistics of $\mathbf{P}$ and $\mathbf{R}$ behavior. We found that in most of these runs $\mathbf{P}$ and $\mathbf{R}$ undergo multiple stochastic switches, in contrast with what is observed in real materials. This indicates that our default values for $Q_P$ and $Q_R$ overestimate thermal fluctuations in monodomain BiFeO$_3$.

Then, we consider a set of $Q_P$ (in the range of 100 to 200~F~m$^{-1}$~s$^{-1}$) and $Q_R$ (in the range of $5\times10^4$ to $8\times10^4$~deg$^2$~m$^3$~J$^{-1}$~s$^{-1}$) combinations and identify optimal choices for which most of our runs ($>90$\%) do not present any spontaneous switching events. The parameter values and switching statistics are summarized in Supplementary Table~S1. 

Next, using optimal $Q_P$ and $Q_R$ pairs, we move to switching simulations under applied field. First, we apply $\mathbf{E}\parallel[00\bar{1}]$ with the amplitude $E_{max}=100$~MV/m, but observe almost no switching events (see detailed statistics in Supplementary Table~S2). Then we increase the field up to $E_{max}=200$~MV/m and find that the most likely switching event is the reversal of $P_z$  as schematically shown in Fig.~\ref{fig:switching_behavior}(a), accompanied by the reversal of $R_z$. An example of the time evolution of $\mathbf{P}$ and $\mathbf{R}$ is shown in Fig.~\ref{fig:switching_mono}, for $Q_P=140$~F~m$^{-1}$~s$^{-1}$ and $Q_R=7\times10^4$~deg$^2$~m$^3$~J$^{-1}$~s$^{-1}$, which give the best statistics of the switching events. More precisely, with these parameters the aforementioned switching path was found in 92.5\% of the considered runs; in 2.5\% there was no switching at all, while in 5\% of the runs the initial reversal of $P_z$ was followed by the reversal of additional $\mathbf{P}$ components resulting from thermal fluctuations in the system. 

\begin{figure*}
    \centering \includegraphics[width=1.0\linewidth,trim=0cm 0cm 0cm 0cm]{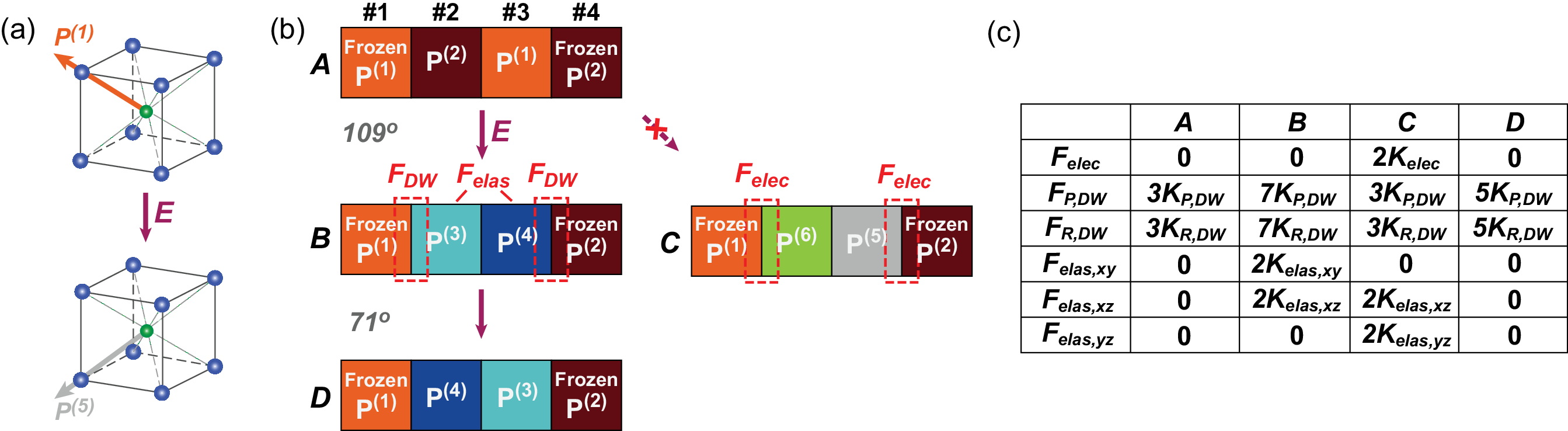}
    \caption{Polarization switching paths that we can expect based on the model for BiFeO$_3$ introduced in Sec.~\ref{subsec:model}. Panel~(a) shows the case of monodomain bulk BiFeO$_3$. Panel (b) shows the case of a multidomain BiFeO$_3$ film, which we approximate by a system of four domains with two of them (\#1 and \#4) frozen, and two (\#2 and \#3) allowed to evolve. Panel (c) shows the contributions from all energy terms introduced in Sec.~\ref{subsec:model} for each of the states (A to D) of panel~(b). For example, state C presents two charged walls -- between domains \#1-\#2 and \#3-\#4, respectively -- and has an associated electrostatic energy penalty given by $F_{elec}$ and labeled ``$2K_{elec}$'' in the Table.}
\label{fig:switching_behavior} 
\end{figure*}

Note that in this simple monodomain situation we expect to observe the switching of only the $z$ component of $\mathbf{P}$ (followed by the $z$ component of $\mathbf{R}$), since $P_z$ is directly coupled to $\mathbf{E}||[00\bar{1}]$, and the reversal of only one component of $\mathbf{P}$ and $\mathbf{R}$ has the lowest energy barrier.
However, as mentioned above, this switching path has never been observed in the experiments of Heron \textit{et al.}~\cite{heron2014}. This indicates that the {\sl multidomain} structure must play a crucial role in determining the experimentally observed switching.

\subsubsection{Polarization switching in a multidomain BiFeO$_3$ film}
\label{subsubsec:multidomain}

Next we study polarization switching in a multidomain BiFeO$_3$ film. As we mentioned in Sec.~\ref{subsec:model}, we approximate it by the one-dimensional series of domains with homogeneous internal structure characterized by $\mathbf{P}_i$, $\mathbf{R}_i$ and $\bm{\eta}_i$ (Fig.~\ref{fig:1dmodel}). We start by considering the system of four domains, two of which (at the ends) are frozen, and two (in the middle) are allowed to evolve in response to applied electric field (Fig.~\ref{fig:switching_behavior}(b)). The frozen domains are necessary for reproducing the state indicated in Fig.~\ref{fig:init_final_domains}(b), where the switching region coexists with parts of the sample that remain in the zero-field configuration. The starting configuration is indicated as the state $A$ in Fig.~\ref{fig:switching_behavior}(b) and presents a pattern of alternating domains $\mathbf{P}^{(1)}$ and $\mathbf{P}^{(2)}$ reflecting the experimental situation (Fig.~\ref{fig:init_final_domains}(a)). This system is described by the full model, Eq.~(\ref{eq:full_model}), introduced in Sec.~\ref{subsec:model}, which includes the couplings related to the presence of the DyScO$_3$ substrate, the formation of domain walls, and the presence of the unswitched region.   

\begin{figure*}
    \centering \includegraphics[width=1.0\linewidth,trim=0cm 0cm 0cm 0cm]{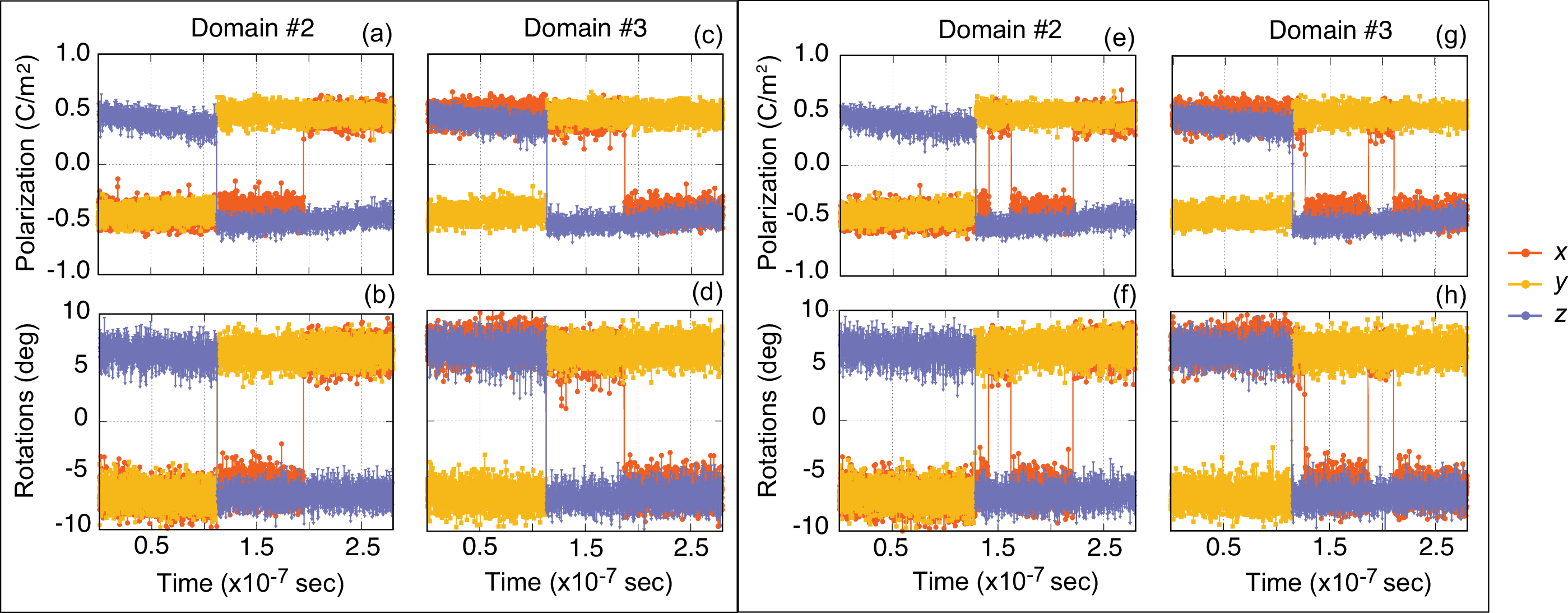}
    \caption{Time evolution of the order parameters in BiFeO$_3$ simulated as a system of four domains shown in Fig.~\ref{fig:switching_behavior}(b) at $T=300$~K under an applied electric field $E_{max}=500$~MV/m using the default parameter set (Table~\ref{tab:default_set}). The applied electric field is time dependent as shown in Fig.~\ref{fig:switching_mono}(c). We show $\mathbf{P}_i$ (top row) and $\mathbf{R}_i$ (bottom row) only for the evolving domains \#2 and \#3. Panels (a)-(d) show the result of one of the runs (out of 200) in which we observed a perfect two-step $109yz+71x$ switch in both evolving domains. Panels (e)-(h) show the result of a different run where the two-step switch in both evolving domains is followed by back-and-forth switches of $P_{i,x}$ and $R_{i,x}$ (see text).}
\label{fig:switching_multi} 
\end{figure*}

First, we discuss the time evolution of $\mathbf{P}_i$ and $\mathbf{R}_i$ as obtained using the default model parameters presented in Table~\ref{tab:default_set}  with no re-scaling ($K=1$) and under an electric field with $E_{max}=500$~MV/m applied along $[00\bar{1}]$ (with smaller fields, $E_{max}=300$ and 400~MV/m, $\mathbf{P}_i$ and $\mathbf{R}_i$ do not switch in most of the runs). The result of one simulation (out of 200) is shown in Figs.~\ref{fig:switching_multi}(a)-(d). One can see that the simultaneous reversal of $P_{i,y}$ and $P_{i,z}$ ($109yz$ switch) in both evolving domains is followed by the reversal of $P_{i,x}$ ($71x$ switch); further, the FeO$_6$ octahedral tilts reverse together with $\mathbf{P}_i$ during this switching process.  
In order to better understand this behavior, let us discuss how the energy terms described in Secs.~\ref{subsubsec:electrostatic}-\ref{subsubsec:elastic_constraints} contribute at each step of the process. 

As we have seen in Sec.~\ref{subsubsec:monodomain}, when $\mathbf{E}$ is applied along $[00\bar{1}]$, one would expect a reversal of only the $z$ component of the polarization, since it is the only one coupled to the field (see state $C$ in Fig. \ref{fig:switching_behavior}(b)). However, if this were to happen, the domain \#2 would form a charged domain wall with the frozen domain \#1, which is strongly penalized by the $F_{elec}$ term in our model. The argument applies also to the domains \#3 and \#4. Therefore, it is preferable to perform a $109yz$ switch resulting in the state $B$ in Fig.~\ref{fig:switching_behavior}(b), which renders neutral domain walls between the switching and fixed domains. However, this state features two $180^\circ$ domain walls (between the domains \#1 and \#2, and between \#3 and \#4) that have a relatively high energy on account of the terms $F_{DW,P}$ and $F_{DW,R}$ (Eqs.~(\ref{eq:F_DW_P}) and (\ref{eq:F_DW_R}), respectively). Additionally, the strain state of the $B$ configuration is penalized by the elastic constraints $F_{sub}$ and $F_{elas,xz}$ (Eqs.~(\ref{eq:F_sub}) and (\ref{eq:F_elas_xz}), respectively). 
Therefore, the energy can be further reduced by performing a  $71x$ switch in both evolving domains, which optimizes all the energy terms in our model. This results in a full reversal of polarization in the domains \#2 and \#3 relative to the initial state $A$.

Note also that the octahedral tilts $\mathbf{R}_i$ follow $\mathbf{P}_i$ at each step, as there is a strong preference for $\mathbf{P}_i$ and $\mathbf{R}_i$ to be perfectly (anti)parallel (see the third term in Eq.~\ref{eq:F_PR}). Further, there is nothing in the energetics of the tilts that tends to favor a different switching path. Hence, the reversal process shown in Figs.~\ref{fig:switching_multi}(a)-(d) is the ideal two-step ferroelectric switching path that permits an accompanying magnetoelectric reversal in BiFeO$_3$.   

It is important to note, however, that for the default parameter set we obtained this ideal behavior in only 5\% of the runs (see Fig.~\ref{fig:statistics_switching_mult}(a)). Additionally, in 6\% of the runs we observed the perfect $109yz+71x$ polarization reversal in only one of the evolving domains, while in the second one $\mathbf{P}_i$ underwent either only a $109yz$ switch or did not switch at all. Nevertheless, the behavior observed in the remaining majority of runs is still physically sound. Indeed, as one can see in Fig.~\ref{fig:statistics_switching_mult}(a), in a large number of the simulations (70\%) we found a $109yz+71x$ $\mathbf{P}_i$ reversal in at least one of the two evolving domains, followed by one or multiple back-switches of $P_{i,x}$. An illustrative example is shown in Figs.~\ref{fig:switching_multi}(e)-(h). 
Here we underline that such in-plane back-switching events are unlikely to occur in real materials, which inevitably contain defects serving as pinning centers and enforcing the system to remain in the lowest-energy state. Pinning, however, is not taken into account in our simulations. For this reason, we assume that switching paths in which a $109yz+71x$ process is followed by a series of $P_x$ back-and-forth switches can be considered as perfect two-step switching events. 
Another observation is that in 21.5\% of the runs we have a $71x$ first switching event in one of the domains. However, this behavior is driven by random noise. Indeed, as we will show in the following, by reducing the noise amplitudes, the number of such switching events can be dramatically reduced (see Fig.~\ref{fig:statistics_switching_mult}(b)). The switching processes indicated in Fig.~\ref{fig:statistics_switching_mult}(a) as "Other" involve, for example, the single-step 180$^\circ$ polarization reversal (with or without $71x$ back-switches of $\mathbf{P}_i$); and $71x+109yz$ (or just $71x$, or just $109yz$) switches in one of the domains with no $\mathbf{P}_i$ switching in the second domain.  

\begingroup
\setlength{\tabcolsep}{9pt} 
\renewcommand{\arraystretch}{1.2} 
\begin{table*}
    \caption{Sets of model parameters giving the highest number of two-step switching events as predicted from our simulations of BiFeO$_3$ approximated by a system of 2 frozen and 2 evolving domains. $K$ is the rescaling factor that we apply to our default choices for $F_L$ and $F_{DW}$. $K_{elas,xz}$ and $K_{elas,yz}$ are the parameters of the elastic constraint terms $F_{sub}$ and $F_{elas,xz}$, respectively. $Q_P$ and $Q_R$ are the noise amplitudes for $\mathbf{P}_i$ and $\mathbf{R}_i$, respectively. $E_{max}$ is the amplitude of the applied electric field. $N_{2d}$ is the number of simulations (in \% out of 200 runs) in which two-step $\mathbf{P}_i$ and $\mathbf{R}_i$ switching occurs in both evolving domains. $N_{1d}$ is the number of simulations (in \% out of 200 runs) in which two-step $\mathbf{P}_i$ and $\mathbf{R}_i$  switching occurs only in one evolving domain, while in the second domain these order parameters switch only by $109^\circ$. $N_{2d}$ and $N_{1d}$ account for the switching events that involve back-and-forth switches of $P_{i,x}$ and $R_{i,x}$.}
\begin{tabular}{ccccccccc}
\hline
\hline
& \centering{i} &  \centering{ii} & \centering{iii} & \centering{iv} & \centering{v} & \centering{vi} & \centering{vii} &  Units
\tabularnewline
\hline
\centering{$K$} & 0.7 & 0.7 & 0.8 & 0.9 & 1.0 & 0.8 & 0.8 & \centering{-}
\tabularnewline
\centering{$K_{elas,xy}$} & 32.043 & 160.218 & 48.065 & 48.065 & 160.218 & 4.600 & 160.218 & \centering{$\times 10^{-18}$, J}
\tabularnewline
\centering{$K_{elas,xz}$} & 32.043 & 160.218 & 48.065 & 48.065 & 160.218 & 4.600 & 160.218 & \centering{$\times 10^{-18}$, J}
\tabularnewline
\hline
\centering{$Q_P$} & 1.0 & 1.0 & 1.4 & 1.4 & 1.8 & 1.2 & 1.2 &  \centering{$\times 10^2$ F m$^{-1}$ s$^{-1}$}
\tabularnewline
\centering{$Q_R$} & 7.0 & 6.0 & 5.0 & 8.0 & 7.0 & 6.0 & 6.0 & \centering{$\times 10^4$ deg$^2$ m$^3$ J$^{-1}$ s$^{-1}$}
\tabularnewline
\hline
\centering{$E_{max}$} & 500 & 600 & 600 & 600 & 600 & 700 & 700 & \centering{MV m$^{-1}$}
\tabularnewline
\hline
\centering{$N_{2d}$} & 57 & 70 & 43 & 55 & 45 & 63 & 68.5 & \%
\tabularnewline
\centering{$N_{1d}$} & 33 & 26 & 44 & 38.5 & 40 & 28.5 & 25.5 & \%
\tabularnewline
\hline
\hline
\end{tabular}
\label{tab:seven_sets}
\end{table*}

\subsection{Parameter optimization}
\begin{figure*}
    \centering \includegraphics[width=0.98\linewidth,trim=0cm 0cm 0cm 0cm]{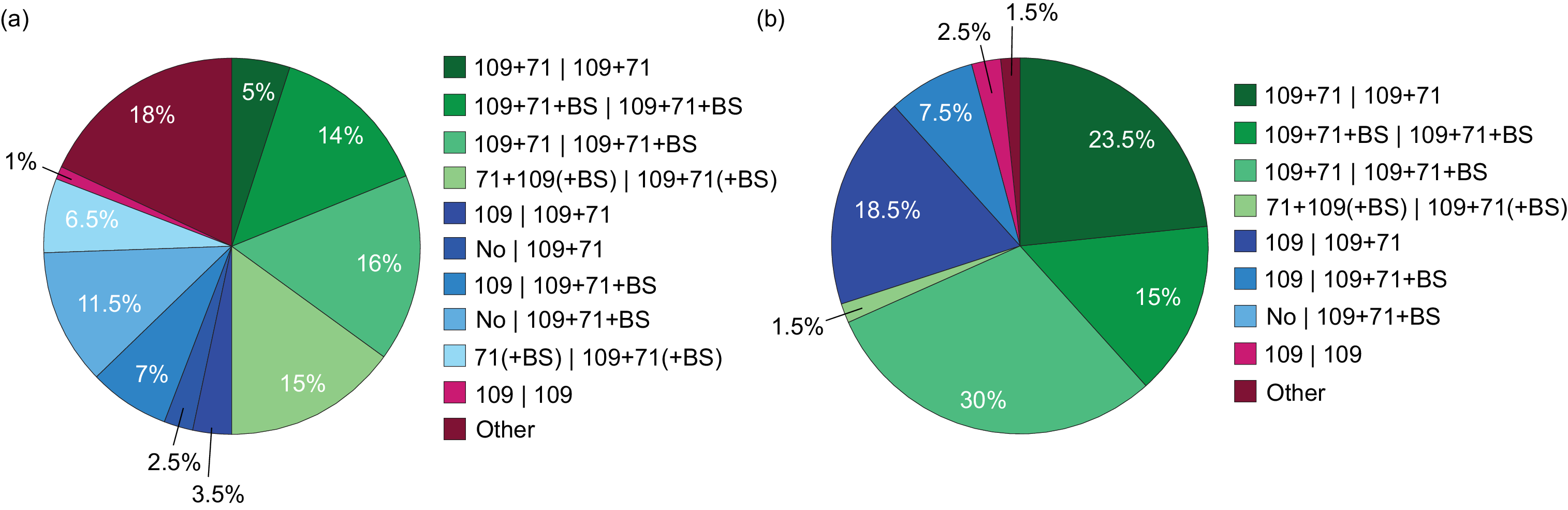}
    \caption{Statistics of polarization switching events in the dynamical simulations of BiFeO$_3$  approximated by a system of 4 domains (2 frozen and 2  allowed to evolve, see Fig.~\ref{fig:switching_behavior}(b)). The charts show the number of simulations (in \% out of 200 runs) in which certain switching paths have been observed. Panel~(a) shows the results obtained using the default parameter set in Table~\ref{tab:default_set}, while panel~(b) shows the results for the parameter set giving the highest number of $109yz+71x$ switches (set~(ii) in Table~\ref{tab:seven_sets}). "+BS" indicates the switching paths that involve back-switches of $P_{i,x}$ and $R_{i,x}$. "+(BS)" indicates that the reported number includes switching events both with and without back-switching.}
\label{fig:statistics_switching_mult}
\end{figure*}

Having introduced the basic results of our simulations, we now describe how the parameters of the model can be adjusted in order to reproduce the experimental observations better.

In Table~\ref{tab:seven_sets} we present seven parameter sets which we select as those giving the switching statistics closest to the experiment. These include the set that allows to use the smallest electric field, labeled (i); the set that does not involve re-scaling of the model parameters ($K=1$), labeled~(v); and the set in which all constants $K_{elas}$ in Eqs.~(\ref{eq:F_sub}), (\ref{eq:F_elas_xz}) and (\ref{eq:F_elas_yz}) are set equal to the elastic constant $C_{44}$ of BiFeO$_3$ obtained from DFT, labeled~(vi). The remaining sets give a high number of two-step switches in 200 runs. In Fig.~\ref{fig:statistics_switching_mult}(c) we show the switching statistics for the representative case of set~(ii), while the corresponding data for the other parameter sets are summarized in Supplementary Fig.~S1. 

One can see that by balancing the noise amplitudes and the heights of the energy barriers in the Landau potential, we can go up to 70\% of two-step switching events in both evolving domains (this number goes down to 23.5\% if we do not correct for spurious back-and-forth switches).  

\begin{figure*}
    \centering \includegraphics[width=1.0\linewidth,trim=0cm 0cm 0cm 0cm]{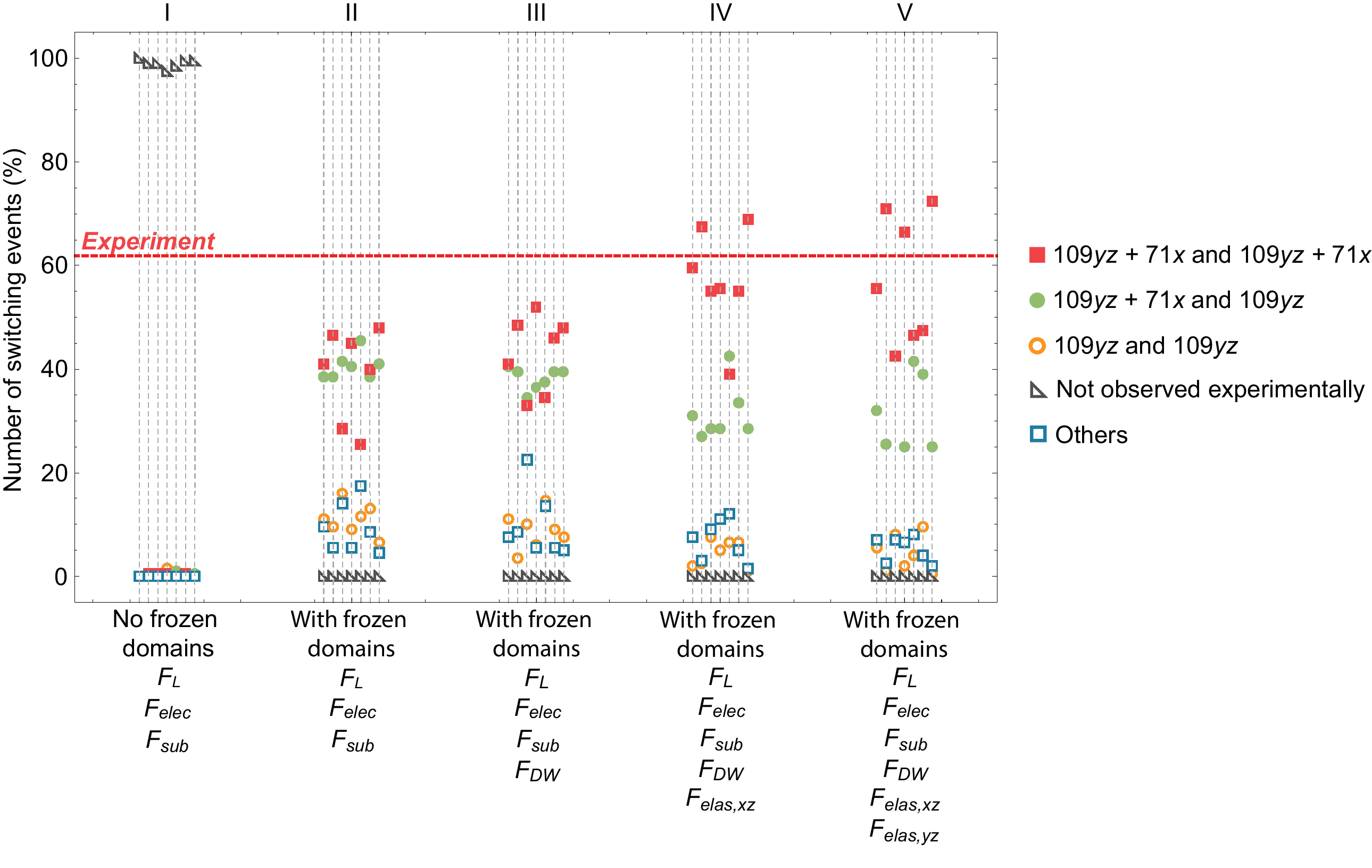}
    \caption{Statistics of switching events for the models and simulation parameters discussed in the text. More precisely, results are presented for five models of increasing complexity. For each model we consider the seven parameter sets presented in Table~\ref{tab:seven_sets}; each parameter set corresponds to a vertical dashed line. Different symbols denote different switching behaviors in domains \#2 and \#3. Note that the switching events indicated as 109$yz$+71$x$ include both perfect two-step switches as well as the paths in which the 109$yz$+71$x$ reversal is followed by stochastic $71x$ back-and-forth jumps. The events denoted as "Not observed experimentally" correspond to $71z$ jumps, or a $109yz$ switch followed by a $71y$ rotation; such events have not been observed in the experiments of Heron \textit {et al}.~\cite{heron2014}. Switching events denoted as "Others" include single-step $180^\circ$ $\mathbf{P}$ switches where $\mathbf{R}$ does not follow, as well as switching paths that start with a $71x$ rotation. The horizontal red dashed line shows the experimentally reported fraction of the sample in which the polarization reversal occurs in two steps~\cite{heron2014}.}
\label{fig:diif_models} 
\end{figure*}

Next, we check that the observed two-step polarization switching behavior is not dependent on the small number of free domains (which is 2) used in the simulations described above. To do that,
we repeat our analysis for systems of 6, 8 and 10 domains (in all cases, the two domains at the ends are frozen) using the parameter sets presented in Table~\ref{tab:seven_sets}.  We observe a high percentage of two-step polarization reversals for all considered system sizes (details in Supplementary Figs.~S2-S4). Therefore, we conclude that our minimal model of 2 free domains captures correctly the two-step polarization reversal.

\subsection{Origin of two-step polarization reversal}

In this section we analyze which terms of the model described by Eq.~(\ref{eq:full_model}) are the key ingredients for driving two-step polarization switching in BiFeO$_3$. For that we consider the system of four domains (two fixed and two free) and perform polarization switching simulations using five models of increasing complexity. Each of the considered models includes the Landau energy of the isolated domains ($F_L$, Eq.~(\ref{eq:Landau})), the electrostatic energy penalty that prevents formation of charged domain walls ($F_{elec}$, Eq.~(\ref{eq:F_elec})) and the elastic energy constraint that accounts for the presence of the DyScO$_3$ substrate ($F_{sub}$, Eq.~(\ref{eq:F_sub})). For every model we consider the seven parameter sets presented in Table~\ref{tab:seven_sets}; we perform 200 runs for each combination of model and parameter set to obtain statistics of the switching events. 

We start with the model that includes only $F_L$, $F_{elec}$ and $F_{sub}$, and perform simulations \textit{without} freezing the domains \#1 and \#4 (see Fig.~\ref{fig:switching_behavior}(b)) while applying periodic boundary conditions (domains \#1 and \#4 are coupled). As one can see in Fig.~\ref{fig:diif_models} (model~I),  we mostly obtain the $71z$ switching path similar to what we find for monodomain BiFeO$_3$ (Figs.~\ref{fig:switching_mono} and  \ref{fig:switching_behavior}(a)). As already mentioned, this switching path was not observed in the experiments of Heron \textit{et al} \cite{heron2014}. 

Then, we freeze the domains \#1 and \#4 and repeat the analysis; the result is shown in Fig.~\ref{fig:diif_models} (model~II). One can see that this immediately gives a high percentage of $109yz+71x$ switching events. There are two main effects at play here. One the one hand, direct $71z$ switching events (which would lead to the state $C$ shown in Fig.~\ref{fig:switching_behavior}(b)) are avoided as those would result in charged domain walls with the fixed (yet-unswitched) regions. Thus we obtain the $109yz$ path instead of $71z$. On the other hand, as explained in Sec.~\ref{subsubsec:multidomain}, the substrate-imposed constraint favors the additional $71x$ jump to further optimize the energy. Hence, remarkably, this very simple model already yields the two-step polarization reversal as the dominant switching path.

Further, the ratio of the $109yz+71x$ events can be increased by adding energy penalties for the structural discontinuity at the domain walls 
$F_{DW}$ (model~III in Fig.~\ref{fig:diif_models}) and remaining elastic constraints (models~IV and V in Fig.~\ref{fig:diif_models}). For the latter, the addition of $F_{elas,xz}$ (model~IV) allows to raise the number of two-step switches up to the experimentally reported ratio of 62\%. By contrast,  $F_{elas,yz}$ does not play a significant role. This makes physical sense, since $F_{elas,yz}$ only penalizes the switching to the domains with $\eta_{yz}>0$ (i.e., $\mathbf{P}^{(5)}$, $\mathbf{P}^{(6)}$, $\mathbf{P}^{(7)}$ and $\mathbf{P}^{(8)}$), which are already precluded by the $K_{elec}$ term.

\section{Discussion}

We have thus found that the occurrence of a two-step polarization reversal relies on three main factors: (1) The presence of a substrate enforcing the formation of the striped patterns of domains with alternating in-plane shear strains; (2) the electrostatic interactions preventing the formation of charged domain walls; (3) The presence of (yet-)unswitched regions which - in combination with point~(2) - effectively preclude the direct switch of the vertical polarization component. Additionally, we find that the elastic interaction between the already switched domains and the yet-unswitched matrix ($F_{elas,xz}$), as well as the energy penalty associated to the domain-wall structural discontinuity ($F_{DW}$), also play a role in making the two-step reversal path as dominant as observed experimentally.   

\subsection{Optimization of two-step polarization reversal}

\begin{figure*}
    \centering \includegraphics[width=1.0\linewidth,trim=0cm 0cm 0cm 0cm]{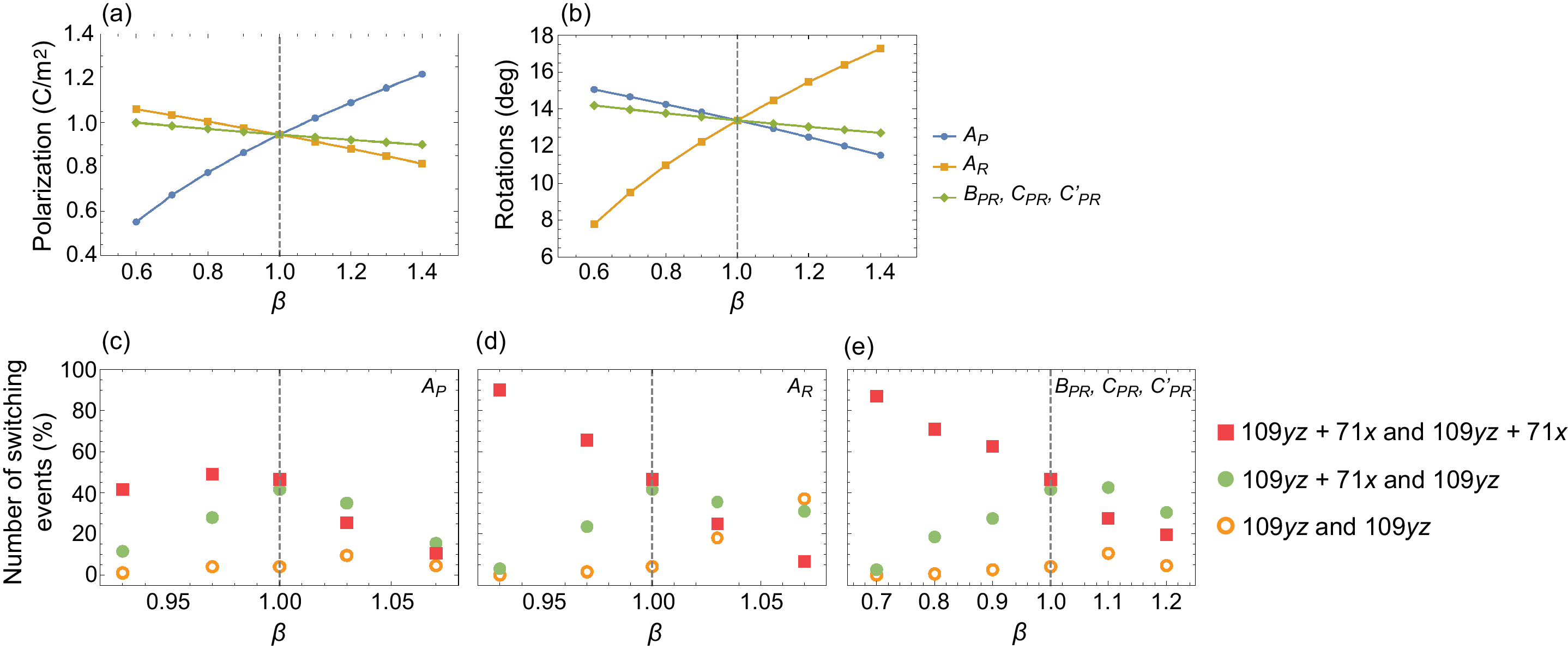}
\caption{Predicted changes as we vary selected parameters of our model (see text). For a given parameter $X$ we consider $X = \beta X^{def}$, where $\beta$ is a scaling factor in the range 0.6 to 1.4, and $X^{def}$ is the value of the considered parameter from the default set. The top row shows the evolution of the magnitudes of electric polarization $\mathbf{P}$ (a) and FeO$_6$ octahedral rotations $\mathbf{R}$ (b) in monodomain BiFeO$_3$. The bottom row shows the number of the switching simulations (in \% out of 200 runs) in which a certain switching path has been observed in the system of four domains. Different symbols denote different switching behaviors in domains \#2 and \#3 (the domains \#1 and \#4 are frozen). Note that the switching events indicated as 109$yz$+71$x$ include both perfect two-step switches as well as  paths in which the 109$yz$+71$x$ reversal is followed by $71x$ back-and-forth jumps. We show the switching statistics as we vary the $A_P$ parameter (c), the $A_R$ parameter (d) and the ${\bf P}-{\bf R}$ coupling parameters (e).}
\label{fig:optimization} 
\end{figure*}

We now use our model to explore potential strategies for optimizing the switching speed and coercive field while preserving (or improving) the ratio of two-step switching events that permit magnetoelectric control of BiFeO$_3$ films. 

These properties of interest can be tuned by modifying the parameters of the Landau energy described by Eq.~(\ref{eq:E_Landau_single}). More precisely, we can control the amplitudes of $\mathbf{P}$ and $\mathbf{R}$, as well as the switching energy barrier heights, by tuning the corresponding quadratic coefficients $A_P$ and $A_R$ of Eqs.~(\ref{eq:F_P}) and (\ref{eq:F_R}). Additionally, we can tune the coupling between $\mathbf{P}$ and $\mathbf{R}$ by modifying the parameters $B_{PR}$, $C_{PR}$ and $C_{PR}'$ of $F(\mathbf{P},\mathbf{R})$ term, Eq. (\ref{eq:F_PR}).  

We start by considering monodomain bulk BiFeO$_3$ (described solely by the Landau term, Eq.~(\ref{eq:E_Landau_single})) and study how changes in the model parameters $A_P$ and $A_R$, as well as $B_{PR}$, $C_{PR}$ and $C_{PR}'$, affect the magnitude of $\mathbf{P}$ and $\mathbf{R}$. First, we use the default parameter values presented in Table~\ref{tab:default_set} and find $\mathbf{P}$ and $\mathbf{R}$ that minimize the energy of the system at $T=0$~K with no applied electric field. Then we consider a series of $A_P$ values, $A_P=\beta A_P^{def}$, where $\beta$ is a number between 0.6 and 1.4, and $A_P^{def}$ is the value of $A_P$ in the default set.
We find the equilibrium $\mathbf{P}$ and $\mathbf{R}$ for each considered $\beta$.
As one can see in Figs.~\ref{fig:optimization}(a) and \ref{fig:optimization}(b), increasing $A_P$ leads to a strong increase in polarization magnitude and small reduction in FeO$_6$ octahedral tilts. Next, we repeat the same exercise, but varying $A_R$. It has similar effect: increasing the magnitude of $A_R$ causes a dramatic increase in the octahedral tilts and a small reduction of the polarization.  Finally, we vary $\beta$ in $B_{PR}=\beta B_{PR}^{def}$, $C_{PR}=\beta C_{PR}^{def}$ and $C_{PR}'=\beta C_{PR}'^{ def}$. As shown in Figs.~\ref{fig:optimization}(a) and \ref{fig:optimization}(b), the simultaneous scaling of all the $\mathbf{P}-\mathbf{R}$ couplings has a small effect on the magnitudes of $\mathbf{P}$ and $\mathbf{R}$: we find that both decrease with increasing $\beta$. 

Next, we analyze how the variation of the model parameters described above affects the switching behavior. For that purpose, we consider our minimal simulated system of four domains, two of which (at the ends) are frozen.  We select the parameter set (v) from Table~\ref{tab:seven_sets} as the starting point, since it does not involve any re-scaling of the Landau energy landscape by the factor $K$ introduced in Sec.~\ref{subsubsec:calibration}.

\begin{figure}
    \centering \includegraphics[width=0.99\linewidth,trim=0cm 0cm 0cm 0cm]{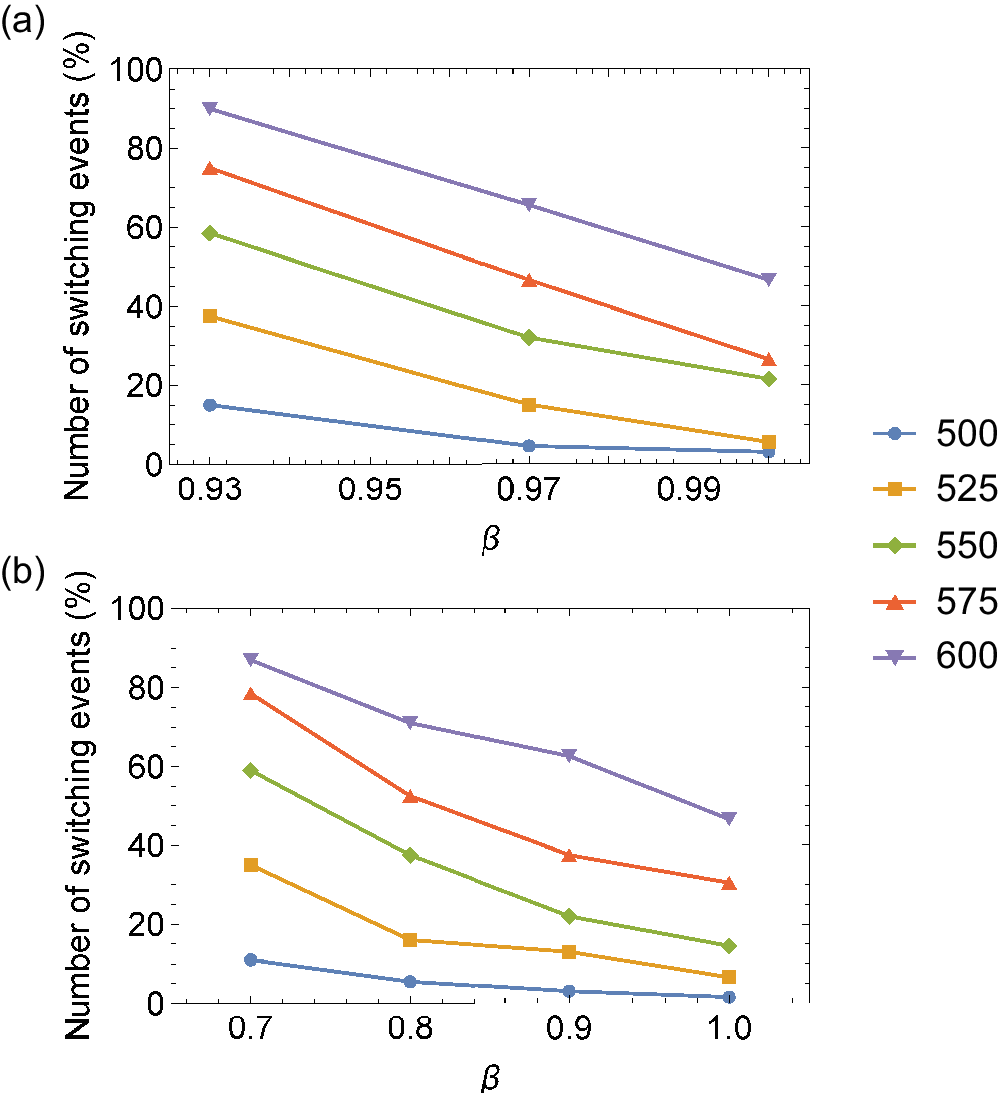}
\caption{Influence of the $A_R$ parameter (a) and ${\bf P}-{\bf R}$ couplings (b) on the ratio of two-step switching events. For a given parameter $X$ we consider $X = \beta X^{def}$, where $\beta$ is a scaling factor and $X^{def}$ is the value of the considered parameter from the default set. We show the number of the switching simulations (in \% out of 200 runs) for the system of 4 domains (2 are frozen and 2 are free) in which $109xy+71x$ switching path have been observed in both evolving domains. The presented values include both perfect two-step switches as well as  paths in which the 109$yz$+71$x$ reversal is followed by back-and-forth $71x$ jumps. The different lines correspond to the different amplitudes of the maximum applied electric field (in MV/m).}
\label{fig:optimization_vs_field} 
\end{figure}

We use this reference model to construct a series of parameter sets that differ only by the value of $A_P$. In particular, we define $A_P$ as $\beta A_P^{def}$ and vary $\beta$ in the range of 0.7 to 1.2.  For each modified set we perform 200 simulations and collect statistics of the switching events. We find that for small $A_P$ ($\beta<0.9$), the local polarizations switch mostly by $180^\circ$ in a single step (which can be also followed by numerous $71x$ back-and-forth jumps). At the same time, in the most of these simulations the octahedral tilts do not switch at all. This makes good physical sense: as we have seen in Figs.~\ref{fig:optimization}(a) and \ref{fig:optimization}(b), small $A_P$ values correspond to small $P_i$ but large $R_i$; the energy barrier for reversing $\mathbf{R}_i$ is likely too high to be overtaken via the coupling to the relatively small polarizations, and it is more favorable to just reverse $\mathbf{P}_i$ in one step (through the $P_i=0$ state) while leaving FeO$_6$ octahedral tilts unaffected. By contrast, for $\beta>1.1$, we observe almost no switching at all, as the energy barrier is too high for the considered amplitudes of thermal noise and applied electric field. Finally, for small changes of $A_P$ ($0.9<\beta<1.1$), we find that it is possible to slightly improve the ratio of two-step switching events (Fig.~\ref{fig:optimization}(c)); yet, the improvement is not significant and we do not discuss it further. 

Next, we repeat the same analysis for $A_R$. As one can see in Fig.~\ref{fig:optimization}(d), a small reduction of $A_R$ yields a significant enhancement of the ratio of two-step switching events. The reason is that the reduction of $A_R$ yields smaller $R_i$ and smaller energy barriers for the switching of $\mathbf{R}_i$; as a result, $\mathbf{R}_i$ follows $\mathbf{P}_i$ more easily during the electric-field-driven switching process. By contrast, an increased $A_R$ leads to a relatively low number of $109yz+71x$ switching events, and the direct $180^\circ$ polarization reversal becomes more favorable. This case is analogous to the one discussed above for a reduced $A_P$, which yields very stiff (unswitchable) tilts.

Thus, we find that a smaller $A_R$ yields an improved ratio of $109yz+71x$ switching events, which suggests that this variation may also enable switching at smaller coercive fields. To check this we repeat the analysis for $\beta \le 1$  using a maximum applied field $E_{max}$ in the range between 500 and 600~MV/m. The obtained statistics of two-step switching events is shown in Fig.~\ref{fig:optimization_vs_field}(a). One can see that reducing $A_R$ indeed allows to obtain high percentages of $109yz+71x$ events at smaller electric fields compared to the default $A_R$. For example, for BiFeO$_{3}$ ($\beta = 1$) we predict that $E_{max} = 600$~MV/m is needed in order to obtain a 46.5\% of two-step switching events; in contrast, for a modified BiFeO$_{3}$ (e.g., a $\beta = 0.93$ in Fig.~\ref{fig:optimization_vs_field}(a)) we obtain the same performance for $E_{max} \lesssim 550$~MV/m, which amounts of a $\sim 10$\% reduction. Therefore, this can be used as an optimization strategy for magnetoelectric switching in BiFeO$_3$. 

Finally, we study the effect of varying the $\mathbf{P}-\mathbf{R}$ coupling. As one can see in Fig.~\ref{fig:optimization}(e), we find that a reduced coupling leads to a dramatic increase in the ratio of $109yz+71x$ switches, since it decreases energy barriers between the states with opposite $\mathbf{P}_i$ and $\mathbf{R}_i$, and allows easier reversal of both degrees of freedom. Similarly to the case of varying $A_R$, in Fig.~\ref{fig:optimization_vs_field}(b) we show the statistics of two-step switching events obtained using $E_{max}$ values from 500 to 600~MV/m. One can see that weakening the $\mathbf{P}-\mathbf{R}$ couplings in BiFeO$_3$ also yields a high number of $109yz+71x$ events at smaller $E_{max}$; therefore, it can serve as an efficient strategy for optimizing switching. 

Based on our findings described above, we can suggest experimental ways to optimize magnetoelectric switching in BiFeO$_3$ films. As we have seen, the largest improvement of the ratio of two-step switching events is achieved by either weakening the FeO$_6$ tilts or by weakening the $\mathbf{P}-\mathbf{R}$ couplings. The former can be achieved experimentally, for example, by creating a solid solution of BiFeO$_3$ with PbTiO$_3$, as the latter has a large polarization, but does not present tilts of TiO$_6$ octahedra. Tuning the coupling between $\mathbf{P}$ and $\mathbf{R}$ might be more challenging, but we can propose some educated guesses. It is well-known that $\mathbf{P}$ in BiFeO$_3$ mainly originates from the displacements of the Bi$^{3+}$ cations, due to the presence of stereochemically active $6s$ lone pairs~\cite{seshadri2001}. Moreover, the large FeO$_6$ octahedral tilts in BiFeO$_3$ are mainly driven by the same kind of chemical mechanism, as the tilts also yield a reduced number of strong Bi-O bonds. Having the same chemical origin, it is not surprising that polarization and octahedral tilts compete strongly in BiFeO$_3$, as they involve alternative ways of satisfying Bi$^{3+}$'s tendency to bond with a reduced number of surrounding oxygens. Therefore, this suggests that weaker $\mathbf{P}-\mathbf{R}$ couplings can be obtained by doping BiFeO$_3$ such that the polarization is less strongly connected to Bi$^{3+}$ and, instead, more dependent on the off-centering of the B-site cation. This logic suggests that a solid solution of BiFeO$_3$ and BaTiO$_3$ (or any other perovskite whose polarization is strongly B-site driven) might have the desired effect.

In summary, we conclude that it should be possible to optimize two-step magnetoelectric switching by doping BiFeO$_3$ in a way that weakens either the octahedral tilts or the coupling between polarization and tilts. Interestingly, solid solutions with well-known perovskites like BaTiO$_3$ and PbTiO$_3$ appear as good candidates for this.

\subsection{Limitations of the model}

As we demonstrated in the previous section, our simple model allows reproducing experimentally observed two-step polarization switching behavior and predicting strategies for its optimization in BiFeO$_3$ thin films. Nevertheless, it is important to bear in mind the drastic approximations our treatment relies on, and its corresponding limitations.

First, the introduced model is not universal. In fact, it
describes a specific case of BiFeO$_3$ film grown in certain conditions (substrate, growth direction and film thickness) \cite{heron2014} and having the striped domain pattern discussed above. Predicting switching behavior in BiFeO$_3$ films having different domain configurations or grown on substrates imposing different elastic constraints would require adapting our simple model in ways that may not be trivial (or possible). 

Second, our model is unable to reproduce the $71x$ polarization rotation being the first step in the switching, though a high number of these events have been reported by Heron \textit{et al}.~\cite{heron2014}. As we briefly mentioned in Sec.~\ref{subsec:experiment}, such $71x$ $\mathbf{P}$ rotations seem related to the movement of the domain walls, which is in turn the result of how neighboring domains rearrange during the transformation. In the simulations described above, each domain was considered as having uniform $\mathbf{P}_i$, $\mathbf{R}_i$ and $\bm{\eta}_i$, and, therefore, they could not capture this behavior. We tried to address this issue by considering larger domains with an inhomogeneous internal structure (see details in Supplementary Sec.~SIV). However, we were still unable to reproduce the experimentally observed fraction of $71x$ polarization rotations as leading switching event.  

Finally, our model is too simple to predict coercive fields accurately, for multiple reasons. First, we impose by hand a non-switching matrix (i.e., frozen domains at the extrema of our simulation box), a hard constraint that should result in an exaggerated coercivity. Additionally, our model does not take into account the presence of defects that usually facilitate switching by providing nucleation centers. Having said this, we think it is justified to use our model and analysis strategy in order to discuss {\sl relative changes} (trends) of coercive fields as we vary the model parameters, elastic constraints, \textit{etc}. We thus restrict our conclusions to such qualitative considerations.

\section{Conclusions}

In summary, in this article we discuss the model that captures the peculiar two-step polarization reversal that enables an electric reversal of the magnetization in BiFeO$_3$ thin films. We demonstrate that the key ingredients that drive this switching process are the presence of a square-like substrate (like DyScO$_3$) that enforces the formation of a striped domain pattern in the films, as well as the electrostatic interactions that prevent the occurrence of charged domain walls during the switching process. We use our model to explore strategies for optimizing the switching, i.e., to reduce the coercive fields while preserving the dominant two-step polarization reversal that enables magnetoelectric control. We conclude that (and explain why) solid solutions of BiFeO$_3$ with simple ferroelectric perovskites like BaTiO$_3$ or PbTiO$_3$ may be the most interesting options to explore.

\section{Acknowledgements}
This work is funded by the the Semiconductor Research Corporation and Intel via contract no. 2018-IN-2865 (N.S.F. and J.\'I.). We also acknowledge the support of the Luxembourg National Research Fund (Grant C21/MS/15799044/FERRODYNAMICS; J.\'I.) and the European Union's Horizon 2020 Research and Innovation Programme (Marie Sk{\l}odowska-Curie grant agreement SCALES - 897614; J.M.M.). D.E.N. worked on this project while affiliated with Intel Corp.

\bibliographystyle{apsrev}
\bibliography{main.bib}

\end{document}